\newif\ifconfver
\confvertrue        
\ifconfver
    \documentclass[10pt,twocolumn,twoside]{IEEEtran}
\else
    \documentclass[11pt,draftcls,onecolumn]{IEEEtran}
\fi

\usepackage{cite}
\usepackage{graphicx}
\usepackage{psfrag}
\usepackage{subfigure}
\usepackage{url}
\usepackage{stfloats}
\usepackage{amsfonts,amssymb,amsmath,amsbsy,bm,paralist,theorem,cite,ifthen,color}
\usepackage{array}
\usepackage{calc}
\usepackage{subfig}
\usepackage{longtable}
\usepackage{multirow}
\usepackage{algorithmic,algorithm}
\usepackage{graphics,booktabs,epsfig,enumerate}

\newtheorem{Lemma}{Lemma}
\newtheorem{Proposition}{Proposition}
\newtheorem{Theorem}{Theorem}
\newtheorem{Definition}{Definition}

\newtheorem{Remark}{Remark}
\theorembodyfont{\rmfamily}

\begin{document}

\def\blue{\color{blue}}
\def\red{\color{red}}
\definecolor{orange}{RGB}{255,107,0}
\def\orange{\color{orange}}
\definecolor{green}{RGB}{0,180,80}
\def\green{\color{green}}

\newcommand\Ab{\mathbf{A}}
\newcommand\Fb{\mathbf{F}}
\newcommand\Pb{\mathbf{P}}
\newcommand\Qb{\mathbf{Q}}
\newcommand\Rb{\mathbf{R}}
\newcommand\Ub{\mathbf{U}}
\newcommand\Vb{\mathbf{V}}
\newcommand\Wb{\mathbf{W}}

\newcommand\Ucal{\mathcal{U}}

\newcommand\ab{\bm{a}}
\newcommand\eb{\bm{e}}
\newcommand\hb{\bm{h}}
\newcommand\nb{\bm{n}}
\newcommand\vb{\bm{v}}
\newcommand\wb{\bm{w}}
\newcommand\xb{\bm{x}}
\newcommand\yb{\bm{y}}

\newcommand\Lambdab{\bm{\Lambda}}

\newcommand\lambdab{\bm{\lambda}}
\newcommand\mub{\bm{\mu}}
\newcommand\nub{\bm{\nu}}

\newcommand\Cplx{\mathbb{C}}
\newcommand\zerob{\mathbf{0}}

\newcommand\tr{\mathrm{Tr}}
\newcommand\rank{\mathrm{rank}}

\markboth{Submitted to IEEE Transactions on Signal Processing, May
2014}{Submitted to IEEE Transactions on Signal Processing, May 2014}

\title{Multicell Coordinated
Beamforming with Rate Outage Constraint--Part II: Efficient
Approximation Algorithms } \ifconfver \else {\linespread{1.1} \rm
\fi

\author{\vspace{0.5cm} Wei-Chiang Li$^\ast$, Tsung-Hui Chang, and Chong-Yung
Chi
\thanks{$^\S$
Part of this work was presented in IEEE ICASSP 2013 \cite{Li2013}
The work is supported by the National Science Council, R.O.C., under
Grants NSC-102-2221-E-007-019-MY3 and NSC 102-2221-E-011-005-MY3.}
\thanks{Wei-Chiang Li and Chong-Yung Chi are with Institute of
Communications Engineering \& Department of Electrical Engineering,
National Tsing Hua University, Hsinchu, Taiwan 30013, R.O.C. E-mail:
weichiangli@gmail.com, cychi@ee.nthu.edu.tw}
\thanks{Tsung-Hui Chang is with Department of Electronic and Computer
Engineering, National Taiwan University of Science and Technology,
Taipei, Taiwan 106, R.O.C. E-mail: tsunghui.chang@ieee.org.}
}



\maketitle

\vspace{-\baselineskip}
\begin{abstract}
This paper studies the coordinated beamforming (CoBF) design for the
multiple-input single-output interference channel, provided that
only channel distribution information is known to the transmitters.
The problem under consideration is a probabilistically constrained
optimization problem which maximizes a predefined system utility
subject to constraints on rate outage probability and power budget
of each transmitter. Our recent analysis has shown that the
outage-constrained CoBF problem is intricately difficult, e.g.,
NP-hard. Therefore, the focus of this paper is on suboptimal but
computationally efficient algorithms. Specifically, by leveraging on
the block successive upper bound minimization (BSUM) method in
optimization, we propose a Gauss-Seidel type algorithm, called
distributed BSUM algorithm, which can handle differentiable,
monotone and concave system utilities. By exploiting a weighted
minimum mean-square error (WMMSE) reformulation, we further propose
a Jocobi-type algorithm, called distributed WMMSE algorithm, which
can optimize the weighted sum rate utility in a fully parallel
manner. To provide a performance benchmark, a relaxed approximation
method based on polyblock outer approximation is also proposed.
Simulation results show that the proposed algorithms are
significantly superior to the existing successive convex
approximation method in both performance and computational
efficiency, and can yield promising approximation performance.
\\\\
\noindent {\bfseries Index terms}$-$ Interference channel,
coordinated beamforming, outage probability, convex optimization.
\ifconfver \else
\\\\
\noindent {\bfseries EDICS}: SAM-BEAM, SPC-CCMC, SPC-INTF, SPC-APPL,
OPT-DOPT \fi
\end{abstract}

%

\ifconfver \else \IEEEpeerreviewmaketitle} \fi

\ifconfver \else \newpage \fi

\vspace{-.25cm}
\section{Introduction}\label{sec:intr}

Coordinated multipoint (CoMP) has been recognized as an effective
approach for interference management in wireless cellular networks
\cite{Lee_2012_ComMag}. There are two main types of cooperation,
namely \textit{MIMO cooperation} and \textit{interference
coordination}, which offer a trade-off between performance gain and
induced overhead on the backhaul network \cite{Gesbert10JSAC}. Via
high-capacity delay-free backhaul, the coordinated base stations
(BSs) for the MIMO cooperation share all the channel state
information (CSI) and users' data, so they perform as a virtual
multiple-antenna BS and high spectrum efficiency can be achieved.
For interference coordination, the BSs only share CSI in order to
jointly design, e.g., power allocation and beamforming strategies,
to mitigate the inter-cell interference. Compared with MIMO
cooperation, the interference coodination requires a relatively
modest amount of backhaul communication \cite{Bjornson11TSP}, and
therefore is still viable when the backhaul capacity is limited. To
study the interference coordination scheme, we consider the commonly
used interference channel (IFC) model \cite{Annapureddy2011}, where
multiple transmitters simultaneously communicate with their
respective receivers over a common frequency band, and hence
interfere with each other.

This paper focuses on the multiple-input single-output (MISO) IFC,
wherein the transmitters are equipped with multiple antennas while
the receivers are equipped with single antenna. Our interest lies in
the \textit{coordinated beamforming} (CoBF) design where the
transmitters cooperate to optimize their beamforming vectors in
order to maximize a network-wide utility function, e.g., the sum
rate, proportional fairness rate, harmonic mean rate, or the
max-min-fairness (MMF) rate. Most of the works in the literature
have assumed that the transmitters have the perfect CSI. Under this
assumption, the MMF CoBF problem has been shown to be
polynomial-time solvable \cite{Liu_11} and efficient algorithms have
been proposed \cite{Liu_11,Cai_2012}. However, for the sum rate,
proportional fairness rate and harmonic mean rate, the utility
maximization CoBF problem is difficult and has been shown NP-hard in
general \cite{Liu_11}. As a result, most of the research efforts
have been made in suboptimal but efficient approximation algorithms;
see, e.g.,
\cite{Liu_11,Zakhour_09,Zhang_Cui_2010,Kim_2011,Shi2011_IteMMSE,Nguyen_11,Weeraddana_2013,Hong2012}
and also \cite{Larsson_08,Larsson_etal2009_mag,Schmidt_09} for game
theoretic approaches. Global optimization algorithms are also
available in
\cite{Jorswieck2010_POA,Utschick2012_POA,Zhang_2012_POA}, but they
are efficient only when the number of users is small.

In practical wireless environments, acquiring accurate users' CSI is
difficult, especially in a mobile network. By contrast, the channel
distribution information (CDI) remains unchanged for a relatively
long period of time, and thus is easier to obtain. However, given
only CDI at the transmitters, the data transmission would suffer
from outage with a nonzero probability, i.e., reliable data
transmission cannot be guaranteed all the time, due to channel
fading. In view of this, the outage-aware CoBF design, which
concerns the probability of rate outage, has attracted extensive
attention recently. For example, the outage balancing CoBF problem
was studied in \cite{Kandukuri02,Tan_2011,Huang_2012_GLOBECOM}, the
outage-constrained power minimization problem was considered in
\cite{Kandukuri02,Ghosh_10}, and the outage-constrained utility
maximization problem was studied in
\cite{Lindblom_11,Park2012,Li_13}. It turns out that the outage
probability constrained CoBF problem is a very difficult
optimization problem. Specifically, it has been shown in
\cite{Li_14_cplx} that the outage balancing problem in
\cite{Huang_2012_GLOBECOM} is in fact NP-hard. Besides, the
outage-constrained CoBF problem \cite{Lindblom_11,Park2012,Li_13} is
NP-hard in general with not only the sum rate but also the MMF rate
(under the MISO setting) \cite{Li_14_cplx}. This implies that
efficient algorithms for high-quality approximate solutions are
indispensable. In \cite{Li_13}, a successive convex approximation
(SCA) algorithm and a distributed SCA (DSCA) algorithm were proposed
to handle the outage-constrained CoBF problem. However, the
computational complexity of the two algorithms is high, hence
preventing them from practical scenarios with a moderate to large
number of users.

In this paper, we propose two efficient distributed CoBF algorithms
for the outage-constrained utility maximization problem, one
referred to as the distributed \textit{block successive upper bound
minimization} (DBSUM) algorithm and the other referred to as the
distributed \textit{weighted minimum mean-square error} (DWMMSE)
algorithm. The DBSUM algorithm is a Gauss-Seidel type algorithm,
derived based on a judicious reformulation of the outage-constrained
problem and application of the BSUM method in
\cite{Razaviyayn_2013BSUM}. The DBSUM algorithm can handle a general
class of monotonic, differentiable concave utilities. On the other
hand, the DWMMSE algorithm is custom-devised for the weighted sum
rate utility, and is a Jocobi-type algorithm so that all the
transmitters can update their respective beamformers in a fully
parallel manner. A common merit of the two algorithms is that the
subproblems to be solved at each iteration are easily implementable,
with problem dimension independent of the number of users. So, the
two algorithms are computationally efficient and scalable with the
size of the network. To provide a benchmark for performance
evaluation of the proposed DBSUM and DWMMSE algorithms, we further
present a constraint relaxation technique for the outage-constrained
CoBF problem. The constraint-relaxed problem is solved by a
\textit{polyblock outer approximation} (POA) algorithm
\cite{Tuy2000_Monotonic} to obtain an upper bound for the optimal
utility value of the original outage-constrained CoBF problem, in
spite of tremendous computation time. We show by computer
simulations that the proposed algorithms significantly outperform
the DSCA algorithm \cite{Li_13} in both performance and
computational efficiency, and exhibit better scalability with
respect to (w.r.t.) the number of users. Moreover, by comparing with
the performance upper bound obtained by the POA algorithm, it can be
corroborated that the proposed algorithms achieve high approximation
accuracy in general.

{\bf Synopsis:} In Section \ref{sec:sys_mod_prob_state}, we present
the system model and problem formulations. The proposed DBSUM
algorithm and DWMMSE algorithm are presented in Section
\ref{sec:general_utility} and Section \ref{sec:WMMSE}, respectively.
In Section \ref{sec:polyblock_outer_apprx}, we present the POA
algorithm which serves as a benchmark performance upper bound for
the two proposed algorithms. Simulation results are then provided in
Section \ref{sec:simulation} to demonstrate the efficacy of the
proposed algorithms. Finally, the conclusions are drawn in Section
\ref{sec:conclusion}.

{\bf Notations:} The set of $n$-dimensional real vectors and complex
vectors are denoted by $\mathbb{R}^n$ and $\mathbb{C}^n$,
respectively. The non-negative real vectors is denoted by
$\mathbb{R}_+^n$. The superscripts `$T$' and `$H$' represent the
matrix transpose and conjugate transpose, respectively. We denote
$\|\cdot\|$ as the vector Euclidean norm. $\mathbf{A}\succeq\zerob$
$(\mathbf{A}\succ\zerob)$ and $\mathbf{a}\succeq\zerob$
$(\mathbf{a}\succ\zerob)$ mean that the matrix $\mathbf{A}$ is
positive semidefinite (definite) and the vector $\mathbf{a}$ is
componentwise nonnegative (positive). We use the expression
$\mathbf{x}\sim\mathcal{CN}(\bm{\mu},\mathbf{Q})$ if $\mathbf{x}$ is
circularly symmetric complex Gaussian distributed with mean
$\bm{\mu}$ and covariance matrix $\mathbf{Q}$. We denote
$\exp(\cdot)$ (or simply $e^{(\cdot)}$) as the exponential function,
while $\ln(\cdot)$ and $\Pr\{\cdot\}$ represent the natural log
function and the probability function, respectively. The principal
eigenvalue of a matrix $\mathbf{A}$ is denoted by
$\lambda_{\max}(\mathbf{A})$. $\{a_{ik}\}$ denotes the set of all
$a_{ik}$ with subscripts $i,k$ covering all the admissible integers
that are defined in the context, and $\{a_{ik}\}_k$ denotes the set
of all $a_{ik}$ with the first subscript equal to $i$. The set
$\{a_{ik}\}_{k{\ne}j}$ is defined by the set $\{a_{ik}\}_k$
excluding $a_{ij}$.

\section{System Model and Problem Statement}\label{sec:sys_mod_prob_state}

Consider a $K$-user MISO IFC where $K$ transmitter-receiver pairs
share a common spectral band. Each transmitter is equipped with
$N_t$ antennas, and all the receivers have single antenna. Assume
that transmit beamforming is used for data transmission.
Specifically, let $\xb_i=\wb_i s_i$ denote the signal intended for
user $i$, where $\wb_i\in \mathbb{C}^{N_t}$ and $s_i\in \mathbb{C}$
are the beamforming vector and the information signal, respectively.
The received signal at receiver $i$ is thus given by
\begin{equation}\label{received signal}
x_i
=\hb_{ii}^H\xb_i+\sum_{k=1,k\neq{i}}^K\hb_{ki}^H\xb_k+n_i,~i=1,\dots,K,
\end{equation}
where $\hb_{ki}\in \mathbb{C}^{N_t}$ denotes the MISO channel from
transmitter $k$ to receiver $i$ and $n_i \in \mathbb{C}$ is the
additive noise at receiver $i$ which has zero mean and variance
$\sigma_i^2>0$. The channels $\hb_{ki}$ are assumed to be complex
Gaussian distributed with zero mean and covariance matrix
$\Qb_{ki}\succeq\zerob$, i.e.,
$\hb_{ki}\sim\mathcal{CN}(\zerob,\Qb_{ki})$, for all
$i,k=1,\ldots,K$. Assume Gaussian signaling, e.g.,
$s_i\sim\mathcal{CN}(0,1)$, and that each receiver $i$ decodes the
information $s_i$ from the received signal with other users'
interference treated as noise (i.e., single user detection). Then,
the instantaneous achievable rate (in bits/sec/Hz) of the $i$th user
is given by
\begin{equation}\label{eq:achievable_rate_CSI}
r_i\left(\{\hb_{ki}\}_k,\{\wb_k\}\right)\!=\!\log_2\left(1\!+\!\frac{\left|\hb_{ii}^H\wb_i\right|^2}{\sum_{k\neq{i}}\left|\hb_{ki}^H\wb_k\right|^2\!+\!\sigma^2_i}\right).
\end{equation}

We assume that only CDI is available at the transmitters; that is,
the transmitters know only the channel covariance matrices
$\Qb_{ik}$, $i,k=1,\dots,K$. Under such circumstances, users might
suffer from transmission outage. Specifically, let $R_i>0$ be the
transmission rate of the $i$th user. The outage event that
$r_i(\{\hb_{ki}\}_{k=1}^K,\{\wb_k\}_{k=1}^K)<R_i$ will occur with a
nonzero probability due to channel fading. Our goal is to optimize
the transmit beamformers $\{\wb_i\}_{i=1}^K$ so that a predefined
system utility, which concerns the system throughput or user
fairness, or considers a proper tradeoff between the two, is
maximized under both transmission outage probability and transmit
power constraints. Mathematically, this can be formulated as the
following outage-constrained CoBF problem:
\begin{subequations}\label{UMX_CDI_Pr}
\begin{align}
\max_{\substack{\wb_i\in\Cplx^{N_t},R_i\ge0,\\i=1,\dots,K}}~&U(R_1,\dots,R_K)\label{UMX_CDI_Pr_a}\\
\text{s.t.}~&\Pr\left\{r_i(\{\hb_{ki}\}_k,\{\wb_k\})<R_i\right\}\le\epsilon_i,\label{UMX_CDI_Pr_b}\\
&\|\wb_i\|^2{\le}P_i,~i=1,\dots,K,\label{UMX_CDI_Pr_c}
\end{align}
\end{subequations}
where $U(R_1,\dots,R_K)$ denotes the system utility of interest,
$P_i>0$ is the power constraint of user $i$, and
$\epsilon_i\in(0,1)$ is the maximal tolerable rate outage
probability for $i=1,\dots,K$. The outage probability constraint
\eqref{UMX_CDI_Pr_b} guarantees that the rate outage probability is
no larger than a specified threshold $\epsilon_i$, which is usually
small, e.g., $\epsilon_i=0.1$. According to
\cite{Kandukuri02,Li_13}, the outage probability in
\eqref{UMX_CDI_Pr_b} has a closed-form expression, and constraint
\eqref{UMX_CDI_Pr_b} can be explicitly expressed as
\begin{equation}\label{eq:outage_constraint_CLSFORM}
\ln\rho_i+\frac{(2^{R_i}\!-\!1)\sigma_i^2}{\wb_i^H\Qb_{ii}\wb_i}+\sum_{k{\ne}i}\ln\left(1\!+\!\frac{(2^{R_i}\!-\!1)\wb_k^H\Qb_{ki}\wb_k}{\wb_i^H\Qb_{ii}\wb_i}\right)\le0,
\end{equation}
where $\rho_i\triangleq1-\epsilon_i$ for $i=1,\dots,K$.

As seen from \eqref{eq:outage_constraint_CLSFORM}, the
outage-constrained CoBF problem \eqref{UMX_CDI_Pr} is in general
nonconvex and appears difficult to deal with. In fact, our recent
complexity analyses in \cite{Li_14_cplx} have shown that problem
\eqref{UMX_CDI_Pr} can be computationally intractable. In
particular, it has been shown in \cite{Li_14_cplx} that problem
\eqref{UMX_CDI_Pr} is NP-hard in general for the weighted sum-rate
utility $U(R_1,\dots,R_K)=\sum_{i=1}^K\alpha_iR_i$, where
$\alpha_i>0$ for $i=1,\dots,K$ are the priority weights of users.
Moreover, for the weighted min-rate (also known as the
max-min-fairness (MMF) rate) utility
$U(R_1,\dots,R_K)=\min_{i\in\{1,\dots,K\}}R_i/\alpha_i$, problem
\eqref{UMX_CDI_Pr} is also NP-hard in general if $N_t\ge2$. Since
maximizing the MMF rate is known polynomial-time solvable under
perfect CSI \cite{Liu_11}, this implies that the outage-constrained
CoBF problem \eqref{UMX_CDI_Pr} is indeed more challenging. In view
of the computational intractability of \eqref{UMX_CDI_Pr}, in the
subsequent Section \ref{sec:general_utility} and Section
\ref{sec:WMMSE}, we propose two algorithms that can efficiently
achieve high-quality approximate solutions to problem
\eqref{UMX_CDI_Pr}.

\section{Outage-Constrained CoBF by Distributed BSUM Algorithm}\label{sec:general_utility}

Let us make the following assumptions on the system utility
$U(\cdot)$. Firstly, $U(\cdot)$ is nondecreasing with respect to
$R_1,\dots,R_K$, respectively, as users always desire to increase
the transmission rate as long as it is possible. Secondly,
$U(\cdot)$ is jointly concave with respect to $R_1,\dots,R_K$, as
concavity enforces user fairness \cite{Mo2000}. These assumptions
are general enough to include some commonly adopted system utilities
such as the weighted sum-rate utility, proportional fairness
utility, harmonic mean utility, and the min-rate (MMF rate) utility
\cite{Liu_11}. Under these assumptions, we show in this section how
the outage-constrained problem \eqref{UMX_CDI_Pr} can be efficiently
handled in a distributed manner by the block successive upper bound
minimization (BSUM) method reported in \cite{Razaviyayn_2013BSUM}.

\vspace{-.1cm}
\subsection{Equivalent Reformulation}\label{subsec:equiv_reformulation}

The key ingredient of the proposed method lies in the following
equivalent reformulation of \eqref{UMX_CDI_Pr}:\vspace{-.1cm}

\begin{Proposition}\label{Prop:equiv_reformulation}
Problem \eqref{UMX_CDI_Pr} is equivalent to the following
problem\vspace{-.15cm}
\begin{subequations}\label{UMX_CDI}
\begin{align}
\max_{\wb_i\in\Cplx^{N_t},~i=1,\dots,K}~&U(R_1(\{\wb_i\}),\dots,R_K(\{\wb_i\}))\label{UMX_CDI_a}\\
\text{s.t.}~&\|\wb_i\|^2{\le}P_i,~i=1,\dots,K,\label{UMX_CDI_b}
\end{align}
\end{subequations}

\vspace{-.2cm}\noindent{where}
\begin{equation}\label{eq:R_def}
R_i(\{\wb_k\})\triangleq\log_2(1+\xi_i(\{\wb_k\}_{k{\ne}i})\wb_i^H\Qb_{ii}\wb_i),
\end{equation}
and $\xi_i(\{\wb_k\}_{k{\ne}i})>0$ is a continuously
differentiable function of $\{\wb_k\}_{k{\ne}i}$ and is a unique solution to the equation
\begin{align}
&\Phi_i(\xi_i,\{\wb_k\}_{k{\ne}i})\triangleq\notag\\
&~~~~\ln\rho_i+\sigma_i^2\xi_i+\sum_{k{\ne}i}\ln(1+(\wb_k^H\Qb_{ki}\wb_k){\cdot}\xi_i)=0,\label{eq:xi_def}
\end{align}

\vspace{-.15cm}\noindent{for} $i=1,\dots,K$.
\end{Proposition}
Proposition \ref{Prop:equiv_reformulation} can be proved by
exploiting the fact that the left-hand side function in
\eqref{eq:outage_constraint_CLSFORM} is monotonic\footnote{Note that
$\Phi_i(\xi_i,\{\wb_k\}_{k{\ne}i})$ is strictly increasing w.r.t.
$\xi_i$. Moreover, since $\Phi_i(0,\{\wb_k\}_{k{\ne}i})=\ln
\rho_i<0$ and
$\Phi_i(\sigma_i^{-2}\ln\rho_i^{-1},\{\wb_k\}_{k{\ne}i})=\sum_{k{\ne}i}\ln\big(1+(\wb_k^H\Qb_{ki}\wb_k)(\sigma_i^{-2}\ln\rho_i^{-1})\big)\ge0$,
the solution of $\Phi_i(\xi_i,\{\wb_k\}_{k{\ne}i})=0$ must be
positive, i.e., $\xi_i(\{\wb_k\}_{k{\ne}i})>0$,
$\forall\{\wb_k\}_{k{\ne}i}$, and can be efficiently obtained by
bisection search.} in $\frac{(2^{R_i}-1)}{\wb_i^H\Qb_{ii}\wb_i}$.
The idea is the same as the one reported in \cite[Lemma
1]{Li_14_cplx} and interested readers may refer to \cite[Appendix
A]{Li_14_cplx} for the detailed proof.

By comparing problem \eqref{UMX_CDI} with problem
\eqref{UMX_CDI_Pr}, one can observe that the rate outage constraints
in \eqref{UMX_CDI_Pr} [and \eqref{eq:outage_constraint_CLSFORM}]
have been judiciously incorporated with the objective function and
it is the function $\xi_i(\{\wb_k\}_{k{\ne}i})$ that implicitly
characterizes the impact of cross-link interference plus noise on
receiver $i$. Indeed, as seen from \eqref{eq:R_def},
$R_i(\{\wb_k\})$ is analogous to the achievable rate of a channel
with channel matrix $\Qb_{ii}$ and interference-plus-noise power
$1/\xi_i(\{\wb_k\}_{k{\ne}i})$. The key advantage of reformulation
\eqref{UMX_CDI} is that the constraint set is separable with respect
to the $K$ beamforming vectors $\wb_1,\ldots,\wb_K$, though the
objective function $U(R_1(\{\wb_i\}),\dots,R_K(\{\wb_i\}))$ is
involved with all $\wb_k$ coupled together. Nevertheless, this type
of problems can be conveniently handled by the BSUM method
\cite{Razaviyayn_2013BSUM} in a distributed and low-complexity
manner, yielding an efficient algorithm for solving the the
outage-constrained CoBF problem \eqref{UMX_CDI}.

\vspace{-.2cm}
\subsection{Brief Reiview of BSUM}\label{subsec:BSUM}

In this subsection, using problem \eqref{UMX_CDI} as an example, we
briefly review the BSUM method in \cite{Razaviyayn_2013BSUM}. For
ease of exposition, let us define
\[
\Ucal(\{\wb_k\}) \triangleq U(R_1(\{\wb_k\}),\dots,R_K(\{\wb_k\})).
\]
The BSUM method \cite{Razaviyayn_2013BSUM} is a
block-coordinate-decent-type (BCD) method \cite{BK:Bertsekas1999}
where the block variables are updated in a round-robin fashion,
i.e., following the Gauss-Seidel update rule. For problem
\eqref{UMX_CDI}, $\wb_1,\ldots,\wb_K$ are the $K$ block variables.
In the $n$th iteration, variable $\wb_i$, where
$i:=(n-1~\mathrm{mod}~K)+1$, is updated by solving the problem
\begin{subequations}\label{BSUM}
\begin{align}
\wb_i^{[n]}=\mathrm{arg}\max_{\wb_i\in\mathbb{C}^{N_t}}~&\bar{\Ucal}^{(i)}(\wb_i\mid\{\wb_k^{[n-1]}\})\\
\text{s.t.}~&\|\wb_i\|^2~{\le}~P_i,
\end{align}
\end{subequations}
where $\{\wb_k^{[n-1]}\}$ denote the beamforming vectors obtained in
the $(n-1)$th iteration, and
$\bar{\Ucal}^{(i)}(\wb_i\mid\{\wb_k^{[n-1]}\})$ is a surrogate
function of $\Ucal(\{\wb_k\})$ given $\{\wb_k^{[n-1]}\}$. The
introduction of the surrogate function
$\bar{\Ucal}^{(i)}(\wb_i\mid\{\wb_k^{[n-1]}\})$ provides extra
flexibility in the algorithm design. In particular, rather than
using the original function $\Ucal(\{\wb_k\})$, one may choose an
advisable $\bar{\Ucal}^{(i)}(\wb_i\mid\{\wb_k^{[n-1]}\})$ that can
either make problem \eqref{BSUM} easily solvable or further lead to
a closed-form solution. Hence, the BSUM method is particularly
useful when the original objective function is intricate and
difficult to optimize, which is the case in problem \eqref{UMX_CDI}
since $\xi_i(\{\wb_k\}_{k{\ne}i})$ are implicit functions without
closed-form expression. It has been shown in
\cite{Razaviyayn_2013BSUM} that the BSUM method performs very well
in several practical signal processing and communication
applications.

Theoretically, the BSUM method has the following convergence
property.\vspace{-.05cm}
\begin{Theorem}\label{thm:BSUM_converge}
{\bf\hspace{-.05cm}\cite[Proposition2, Theorem
2(b)]{Razaviyayn_2013BSUM}:} The iterates
$(\wb_1^{[n]},\dots,\wb_K^{[n]})$ converge to the set of stationary
points of problem \eqref{UMX_CDI} as long as
\begin{subequations}\label{eq:regularity}
\begin{align}
&\Ucal(\{\wb_k\})~ \text{is differentiable
in}~ \{\wb_k\}; \label{eq:regularity_e} \\
&\bar{\Ucal}^{(i)}(\wb_i\mid\{\bar{\wb}_k\})\le\Ucal(\wb_i,\{\bar{\wb}_k\}_{k{\ne}i});\label{eq:regularity_a}\\
&\bar{\Ucal}^{(i)}(\bar{\wb}_i\mid\{\bar{\wb}_k\})=\Ucal(\{\bar{\wb}_k\});\label{eq:regularity_b}\\
&\bar{\Ucal}^{(i)}(\wb_i\mid\{\bar{\wb}_k\})~\text{is continuous
in}~(\wb_i,\{\bar{\wb}_k\});\label{eq:regularity_d}\\
&\text{problem \eqref{BSUM} has a unique
solution,}\label{eq:regularity_f}
\end{align}
\end{subequations}
for all $\|\wb_i\|^2\le{P_i}$, $\|\bar{\wb}_k\|^2\le{P_k}$,
$i,k=1,\dots,K$, and $n\ge1$.
\end{Theorem}
Condition \eqref{eq:regularity_e} requires that the system utility
function $U(R_1,\dots,R_K)$ is differentiable, e.g., the weighted
sum-rate utility, the proportional fairness utility and the harmonic
mean utility\footnote{We should mention that the BSUM method
\cite{Razaviyayn_2013BSUM} can also handle non-differentiable
problems, but it requires additional regularity assumption on the
objective function. The non-differentiable MMF rate utility
$U(R_1,\dots,R_K)=\min_{i\in\{1,\dots,K\}}R_i/\alpha_i$
unfortunately does not satisfy the regularity assumption.
Alternative approach to handling the MMF rate utility problem will
be discussed in Section \ref{subsec:max-min}.}. Conditions
\eqref{eq:regularity_a} and \eqref{eq:regularity_b} imply that
$\bar{\Ucal}^{(i)}(\wb_i\mid\{\bar{\wb}_k\})$ is a universal lower
bound of $\Ucal(\wb_i,\{\bar{\wb}_k\}_{k{\ne}i})$ and it is tight
locally when $\wb_i=\bar\wb_i$.

If all the $K$ beamforming vectors are treated as one block variable
$(\wb_1,\ldots,\wb_K)$, then the BSUM method reduces to the
successive upper bound minimization (SUM) method
\cite{Razaviyayn_2013BSUM}. In Section \ref{sec:WMMSE}, we will use
this SUM method to devise another algorithm for problem
\eqref{UMX_CDI} with the weighted sum rate utility.

\subsection{DBSUM for Problem \eqref{UMX_CDI}}\label{subsec:BSUM2}

As seen, to apply the BSUM method to our problem \eqref{UMX_CDI},
one of the key steps is to construct appropriate surrogate functions
$\bar{\Ucal}^{(i)}(\wb_i\mid\{\bar{\wb}_k\})$, $i=1,\ldots,K$, that
satisfy conditions in
\eqref{eq:regularity_a}-\eqref{eq:regularity_f}. It turns out that
this is not a trivial task since there is no explicit expression for
$\xi_i(\{\wb_k\}_{k{\ne}i})$. To overcome this, we notice that, in
\eqref{eq:R_def}, $R_i(\{\wb_k\})$ has some nice monotonicity and
concavity (resp. convexity) with respect to $\wb_i^H\Qb_{ii}\wb_i$
(resp. $\wb_k^H\Qb_{ki}\wb_k$), as stated in the following
lemma.\vspace{-.05cm}

\begin{Lemma}\label{lemma:rate_func_convexity}
For each $i\in\{1,\dots,K\}$, the function $R_i(\{\wb_k\})$ in
\eqref{eq:R_def} is strictly increasing and strictly concave with
respect to $\wb_i^H\Qb_{ii}\wb_i$, while it is nonincreasing and
convex with respect to each $\wb_k^H\Qb_{ki}\wb_k$ where
$k\in\{1,\dots,K\}\setminus \{i\}$.
\end{Lemma}

The proof is given in Appendix \ref{sec:proof_convexity}. Based on
Lemma \ref{lemma:rate_func_convexity}, we propose the following
surrogate function for updating $\wb_i$.
\begin{align}
&\bar{\Ucal}^{(i)}(\wb_i\mid\{\bar{\wb}_k\})\triangleq\notag\\
&~U\!\!\left(\!\bar{R}_1^{(i)}\!(\wb_i|\{\bar{\wb}_k\}\!),\dots,\bar{R}_K^{(i)}\!(\wb_i|\{\bar{\wb}_k\}\!)\!\right)\!-\!\frac{c}{2}\|\wb_i\!-\!\bar{\wb}_i\|^2,\label{def:barUcal}
\end{align}
where $c>0$ is a penalty parameter and\vspace{-.1cm}
\begin{align}
&\bar{R}_j^{(i)}(\wb_i\mid\{\bar{\wb}_k\})\triangleq\label{def:barR}\\
&\begin{cases}
 \log_2\!\big(1\!\!+\!\xi_i(\{\bar{\wb}_k\}_{k{\ne}i})(2\Re\{\bar{\wb}_i^H\Qb_{ii}\wb_i\}\!-\!\bar{\wb}_i^H\Qb_{ii}\bar{\wb}_i)\big),~j\!=\!i,\\
 R_j(\{\bar{\wb}_k\})+\frac{\partial{R}_j(\{\bar{\wb}_k\})}{\partial\wb_i^H\Qb_{ij}\wb_i}\big(\wb_i^H\Qb_{ij}\wb_i-\bar{\wb}_i^H\Qb_{ij}\bar{\wb}_i\big),~j\ne{i},
 \end{cases}\notag
\end{align}

\vspace{-.1cm}\noindent{where} $\Re(x)$ denotes the real part of
$x\in\mathbb{C}$. Since $\wb_i^H\Qb_{ii}\wb_i$ is convex in $\wb_i$,
its first-order approximation w.r.t. $\wb_i=\bar{\wb}_i$
satisfies\vspace{-.1cm}
\begin{align*}
\wb_i^H\Qb_{ii}\wb_i \geq \bar{\wb}_i^H\Qb_{ii}\wb_i+
\wb_i^H\Qb_{ii}\bar{\wb}_i-\bar{\wb}_i^H\Qb_{ii}\bar{\wb}_i,
\end{align*}

\vspace{-.1cm}\noindent{which} implies that
$\bar{R}_i^{(i)}(\wb_i\mid\{\bar{\wb}_k\})$ in \eqref{def:barR}
satisfies\vspace{-.1cm}
\begin{subequations}\label{eq:localtight i}
\begin{align}
  &\bar{R}_i^{(i)}(\wb_i\mid\{\bar{\wb}_k\}) \leq R_i(\wb_i,\{\bar \wb_k\}_{k\neq i}),~\forall \wb_i, \\
  &\bar{R}_i^{(i)}(\bar\wb_i\mid\{\bar{\wb}_k\}) = R_i(\{\bar \wb_k\}).
\end{align}
\end{subequations}

\vspace{-.1cm}\noindent{Moreover}, it is clear that
$\bar{R}_i^{(i)}(\wb_i\mid\{\bar{\wb}_k\})$ is concave in $\wb_i$.

For $j\neq i$, since $R_j(\wb_i,\{\bar \wb_k\}_{k\neq i})$ is convex
w.r.t. $\wb_i^H\Qb_{ij}\wb_i$ according to Lemma
\ref{lemma:rate_func_convexity}, its first-order approximation
w.r.t. $\wb_i^H\Qb_{ij}\wb_i=\bar{\wb}_i^H\Qb_{ij}\bar{\wb}_i$,
i.e., $\bar{R}_j^{(i)}(\wb_i\mid\{\bar{\wb}_k\})$ in
\eqref{def:barR} for $j\neq i$, satisfies\vspace{-.1cm}
\begin{subequations}\label{eq:localtight j}
\begin{align}
  &\bar{R}_j^{(i)}(\wb_i\mid\{\bar{\wb}_k\}) \leq R_j(\wb_i,\{\bar \wb_k\}_{k\neq i}),~\forall \wb_i, \\
  &\bar{R}_j^{(i)}(\bar\wb_i\mid\{\bar{\wb}_k\}) = R_j(\{\bar \wb_k\}).
\end{align}
\end{subequations}

\vspace{-.1cm}\noindent{A} closed-form expression of the partial
derivative
$\frac{\partial{R}_j(\{\bar{\wb}_k\})}{\partial\wb_i^H\Qb_{ij}\wb_i}$
in \eqref{def:barR} is given on the top of the next page,
\begin{figure*}[t]
\begin{align}
&\frac{\partial{R}_j(\{\bar{\wb}_k\})}{\partial\wb_i^H\Qb_{ij}\wb_i}=
\left.\frac{\partial\log_2(1+\xi_j\cdot\bar{\wb}_j^H\Qb_{jj}\bar{\wb}_j)}{\partial\xi_j}
\right|_{\xi_j=\xi_j(\{\bar{\wb}_k\}_{k{\ne}j})}\times\left.\frac{\partial\xi_j(\{\wb_k\}_{k{\ne}j})}{\partial\wb_i^H\Qb_{ij}\wb_i}\right|_{\wb_k=\bar{\wb}_k,\forall{k{\ne}i}}\notag\\
&~~=\frac{\bar{I}_{jj}}{\ln2\cdot(1+\xi_j(\{\bar{\wb}_k\}_{k{\ne}j})\bar{I}_{jj})}\times\frac{-\xi_j(\{\bar{\wb}_k\}_{k{\ne}j})}{1+\bar{I}_{ij}\xi_j(\{\bar{\wb}_k\}_{k{\ne}j})}\frac{1}{\sigma_j^2+\sum_{\ell{\ne}j}\bar{I}_{\ell{j}}(1+\bar{I}_{\ell{j}}\xi_j(\{\bar{\wb}_k\}_{k{\ne}j}))^{-1}}\notag\\
&~~=\frac{-\bar{I}_{jj}\xi_j(\{\bar{\wb}_k\}_{k{\ne}j})}{\ln2\cdot(1+\xi_j(\{\bar{\wb}_k\}_{k{\ne}j})\bar{I}_{jj})}\bigg[(1+\bar{I}_{ij}\xi_j(\{\bar{\wb}_k\}_{k{\ne}j}))\cdot\bigg(\sigma_j^2+\sum_{\ell{\ne}j}\frac{\bar{I}_{\ell{j}}}{1+\bar{I}_{\ell{j}}\xi_j(\{\bar{\wb}_k\}_{k{\ne}j})}\bigg)\bigg]^{-1}\le0\label{eq:partial_derivative}
\end{align}

\vspace{-.1cm}\hrulefill\vspace{-.3cm}
\end{figure*}
where $\bar{I}_{kj}\triangleq\bar{\wb}_k^H\Qb_{kj}\bar{\wb}_k$ for
all $j,k$, and the second equality is obtained by applying the
implicit function theorem \cite{Krantz_Parks02} (for computing
$\frac{\partial\xi_j(\{\wb_k\}_{k{\ne}j})}{\partial\wb_i^H\Qb_{ij}\wb_i}$).
Since
$\frac{\partial{R}_j(\{\bar{\wb}_k\})}{\partial\wb_i^H\Qb_{ij}\wb_i}$
is non-positive (see \eqref{eq:partial_derivative}),
$\bar{R}_j^{(i)}(\wb_i|\{\bar{\wb}_k\})$ in \eqref{def:barR} for
$j\neq i$ is concave in $\wb_i$. Besides, by the fact that
$\xi_i(\cdot)$ is a continuously differentiable function (see
Proposition \ref{Prop:equiv_reformulation}),
$\bar{R}_j^{(i)}(\wb_i\mid\{\bar{\wb}_k\})$ is continuous in
$(\wb_i,\{\bar{\wb}_k\})$, for all $j=1,\ldots,K$.

The surrogate function $\bar{\Ucal}^{(i)}(\wb_i\mid\{\bar{\wb}_k\})$
in \eqref{def:barUcal} thereby has the following properties. First,
from \eqref{eq:localtight i}, \eqref{eq:localtight j}, continuity of
$\bar{R}_j^{(i)}(\wb_i\mid\{\bar{\wb}_k\})~\forall j$, and the
monotonicity of $U(R_1,\ldots,R_K)$, we conclude that
$\bar{\Ucal}^{(i)}(\wb_i\mid\{\bar{\wb}_k\})$ in \eqref{def:barUcal}
satisfies the conditions
\eqref{eq:regularity_a}-\eqref{eq:regularity_d}. Second, from the
concavity of $\bar{R}_j^{(i)}(\wb_i\mid\{\bar{\wb}_k\})$,
$j=1,\ldots,K$, monotonicity and concavity of $U(R_1,\ldots,R_K)$
and the quadratic penalty $-\frac{c}{2}\|\wb_i-\bar{\wb}_i\|^2$, the
surrogate function $\bar\Ucal^{(i)}(\wb_i \mid \{\bar\wb_k\})$ in
\eqref{def:barUcal} is strongly concave, which infers that
\eqref{eq:regularity_f} holds true. Therefore, we conclude that the
BSUM method in \eqref{BSUM} and \eqref{def:barUcal} has the
following convergence property:

\vspace{-.2cm}
\begin{Proposition}\label{prop:BSUM_converge}
Suppose that the system utility $U(R_1,\dots,R_K)$ is
differentiable, jointly concave, and is nondecreasing w.r.t. each
$R_i$, $i=1,\dots,K$. Then, the sequence
$\big\{\Ucal\big(\{\wb_k^{[0]}\}\big),\Ucal\big(\{\wb_k^{[1]}\}\big),\ldots\big\}$
generated by the BSUM method in \eqref{BSUM} and \eqref{def:barUcal}
converges monotonically, and every limit point of the sequence
$\big\{\big(\wb_1^{[n]},\dots,\wb_K^{[n]}\big)\big\}_{n=1}^\infty$
is a stationary point of problem \eqref{UMX_CDI}.
\end{Proposition}

\vspace{-.1cm}\emph{Proof:} Let $i:=(n-1~\mathrm{mod}~K)+1$. Then we
have
\begin{align}\label{eq: bsum monotone}
\Ucal\big(\{\wb_k^{[n]}\}\big)&=\Ucal\big(\wb_i^{[n]},\{\wb_k^{[n-1]}\}_{k{\ne}i}\big)\notag\\
&\ge\bar{\Ucal}^{(i)}\big(\wb_i^{[n]}\mid\{\wb_k^{[n-1]}\}\big)\notag\\
&\ge\bar{\Ucal}^{(i)}\big(\wb_i^{[n-1]}\mid\{\wb_k^{[n-1]}\}\big)\notag\\
&=\Ucal\big(\{\wb_k^{[n-1]}\}\big),
\end{align}
where the first inequality comes from \eqref{eq:regularity_a}, the
second inequality comes from the optimality of $\wb_i^{[n]}$ to
problem \eqref{BSUM}, and the last equality results from
\eqref{eq:regularity_b}. Equation \eqref{eq: bsum monotone} implies
that the system utility is nondecreasing from one iteration to
another. On the other hand, due to the transmit power constraints
\eqref{UMX_CDI_b}, the sequence
$\{\Ucal\big(\{\wb_k^{[n]}\}\big)\big\}_{n=1}^\infty$ is bounded.
Hence, the system utility converges monotonically. As previously
mentioned, the surrogate functions
$\bar{\Ucal}^{(i)}(\wb_i\mid\{\bar{\wb}_k\})$ satisfies the
conditions in \eqref{eq:regularity}. Therefore, we obtain from
Theorem \ref{thm:BSUM_converge} that every limit point of the
sequence
$\{\big(\wb_1^{[0]},\dots,\wb_K^{[0]}\big),\big(\wb_1^{[1]},\dots,\wb_K^{[1]}\big),\ldots\}$
is a stationary point of problem
\eqref{UMX_CDI}.\hfill{$\blacksquare$}

As $\bar\Ucal^{(i)}(\wb_i \mid \{\bar\wb_k\})$ in
\eqref{def:barUcal} is (strongly) concave, problem \eqref{BSUM} is a
convex problem which is efficiently solvable. More importantly, the
BSUM method can be implemented in a distributed manner, as only one
user is involved at each iteration. Information required for solving
\eqref{BSUM} can be obtained through message exchange between users.
This leads to the proposed DBSUM algorithm as detailed in Algorithm
\ref{alg:BSUM_UMX}.

\vspace{-.08cm}
\begin{algorithm}[h]
\caption{DBSUM algorithm for handling problem
\eqref{UMX_CDI}}
\begin{algorithmic}[1]\label{alg:BSUM_UMX}
\STATE {\bf Given} a set of beamformers $\{\wb_i^{[0]}\}$ satisfying
\eqref{UMX_CDI_b}, and set $n:=0$; Transmitter $i$ sends the
quantity $(\wb_i^{[0]})^H\Qb_{ij}\wb_i^{[0]}$ to transmitter $j$,
$\forall{j\neq{i}}$, $i=1,\dots,K$.

\REPEAT

\STATE $n:=n+1$;

\STATE $i:=(n-1~\mathrm{mod}~K)+1$;

\STATE For all $j{\ne}i$, transmitter $j$ computes
$R_j(\{\wb_k^{[n-1]}\})$ and
$\frac{\partial{R}_j(\{\wb_k^{[n-1]}\})}{\partial\wb_i^H\Qb_{ij}\wb_i}$
by \eqref{eq:R_def} and \eqref{eq:partial_derivative}, respectively,
and sends them to transmitter $i$;

\STATE Transmitter $i$ solves \eqref{BSUM} using \eqref{def:barUcal}
and \eqref{def:barR} to obtain $\wb_i^{[n]}$, and sends the quantity
$(\wb_i^{[n]})^H\Qb_{ij}\wb_i^{[n]}$ to transmitter $j$,
$\forall{j\ne{i}}$;

\STATE $\wb_k^{[n]}:=\wb_k^{[n-1]}$, $\forall{k\ne{i}}$;

\UNTIL the predefined stopping criterion is met.

\STATE {\bf Output} $\{\!\wb_i^{[n]}\!\}$ as an approximate solution
of problem \eqref{UMX_CDI}.
\end{algorithmic}
\end{algorithm}

\vspace{-.3cm}
\subsection{MMF Rate Utility Maximization}\label{subsec:max-min}

Unfortunately, the MMF rate utility
$U(R_1,\dots,R_K)=\min_{i\in\{1,\dots,K\}}R_i/\alpha_i$ is not
differentiable, and thus the DBSUM algorithm (Algorithm
\ref{alg:BSUM_UMX}) cannot directly be applied. To resolve this
issue, we consider the \textit{log-sum-exp} approximation of the
$\min$ function \cite{BK:BoydV04}; specifically, it is known that
\begin{equation}\label{eq:log_sum_exp_apprx}
\min_{n\in\{1,\dots,N\}}\!\!a_n\!\ge\!-\frac{1}{\gamma}\log_2\!\!\left(\sum_{n=1}^N2^{-{\gamma}a_n}\!\!\right)\!\ge\!\min_{n\in\{1,\dots,N\}}\!\!a_n-\frac{\log_2\!N}{\gamma}
\end{equation}
where $\gamma$ can be any positive real value. The inequalities in
\eqref{eq:log_sum_exp_apprx} show that
$-\frac{1}{\gamma}\log_2\left(\sum_{n=1}^N2^{-\gamma{a_n}}\right)$
can be used as an approximation of $\min_{n\in\{1,\dots,N\}}a_n$,
and the approximation error is no larger than
$\frac{\log_2N}{\gamma}$. By \eqref{eq:log_sum_exp_apprx}, we
approximate the MMF rate utility as\vspace{-.08cm}
\[
\min_{i\in\{1,\dots,K\}}\frac{R_i}{\alpha_i}\!\approx\!-\frac{1}{\gamma}\log_2\!\!\left(\sum_{i=1}^K2^{-\gamma{R}_i/\alpha_i}\!\right)\!\triangleq\!\bar{U}(R_1,\dots,R_K).
\]

\vspace{-.08cm}\noindent{It} is readily to see that
$\bar{U}(R_1,\dots,R_K)$ is differentiable, jointly concave in
$(R_1,\dots,R_K)$, and is strictly increasing w.r.t. each $R_i$,
$i=1,\dots,K$. Therefore, the DBSUM algorithm can be applied.

\vspace{-.1cm}
\begin{Remark}{\rm
The DSCA algorithm proposed in \cite{Li_13} handles the
outage-constrained problem \eqref{UMX_CDI_Pr} in a similar fashion
as the proposed DBSUM algorithm, but the former solves a more
involved subproblem \cite[Eqn. (36)]{Li_13} than the latter at each
iteration. Specifically, the problem size, i.e., number of variables
and number of constraints, of \cite[Eqn. (36)]{Li_13} is in the
order of $K$. By contrast, the problem size of the subproblem
\eqref{BSUM} in Algorithm \ref{alg:BSUM_UMX} is independent of $K$.
Moreover, problem \eqref{BSUM} has a simple 2-norm constraint, which
makes it easily implementable by using, e.g., the gradient
projection method \cite[Section 2.3.1]{BK:Bertsekas1999}. We will
show by simulations that Algorithm \ref{alg:BSUM_UMX} is indeed
computationally more efficient than the DSCA algorithm. }
\end{Remark}

\section{Distributed WMMSE Algorithm for Weighted Sum Rate Maximization}\label{sec:WMMSE}

In the previous section, the DBSUM algorithm for problem
\eqref{UMX_CDI} updates the beamforming vectors in the Gauss-Seidel
manner, though it can handle a general utility function. In this
section, we focus on the weighted sum rate (WSR) utility
$U(R_1,\dots,R_K)=\sum_{i=1}^K\alpha_iR_i$ and further propose a
Jacobi-type distributed algorithm where the beamforming vectors are
updated in parallel at each iteration. The idea behind is a
judicious combination of the SUM method (i.e., the BSUM method with
only one block) \cite{Razaviyayn_2013BSUM} and the WMMSE
reformulation \cite{Shi2011_IteMMSE}. To proceed, let us rewrite
\eqref{UMX_CDI} with the WSR utility here
\begin{subequations}\label{WSRM}
\begin{align}
\max_{\wb_i\in\mathbb{C}^{N_t},~i=1,\dots,K}~&\Ucal_{wsr}(\{\wb_k\})\label{WSRM_a}\\
\text{s.t.}~&\|\wb_i\|^2{\le}P_i,~i=1,\dots,K,\label{WSRM_b}
\end{align}
\end{subequations}
where
$\Ucal_{wsr}\!(\{\wb_k\})\!\!\triangleq\!\!\sum_{i=1}^K\!\alpha_i\!\log_2(1\!+\!\xi_i(\{\wb_k\}_{k{\ne}i}\!)\wb_i^H\Qb_{ii}\wb_i)$.

We aim to handle \eqref{WSRM} by the SUM method
\cite{Razaviyayn_2013BSUM}, using a properly designed surrogate
function that is amenable to parallel implementation. To the end,
let us recall the function $\Phi_i(\zeta_i,\{\wb_k\}_{k{\ne}i})$ in
\eqref{eq:xi_def}. Given any feasible point $\{\bar{\wb}_k\}$
satisfying \eqref{WSRM_b}, $\Phi_i(\zeta_i,\{\wb_k\}_{k{\ne}i})$ has
an upper bound as follows\vspace{-.1cm}
\begin{align}
&\Phi_i(\zeta_i,\{\wb_k\}_{k{\ne}i})\!=\!\ln\rho_i\!+\!\sigma_i^2\zeta_i\!+\!\sum_{k{\ne}i}\ln(1\!+\!\wb_k^H\Qb_{ki}\wb_k\zeta_i)\notag\\
&\le\ln\rho_i+\sigma_i^2\zeta_i+\sum_{k\ne{i}}\ln(1\!+\!\bar{\wb}_k^H\!\Qb_{ki}\bar{\wb}_k\bar{\zeta}_i)\notag\\
&~~~~~+\sum_{k{\ne}i}\frac{\wb_k^H\!\Qb_{ki}\wb_k\zeta_i\!-\!\bar{\wb}_k^H\!\Qb_{ki}\bar{\wb}_k\bar{\zeta}_i}{1+\bar{\wb}_k^H\Qb_{ki}\bar{\wb}_k\bar{\zeta_i}}\notag\\
&=\ln\rho_i\!+\!\sum_{k{\ne}i}\ln(1\!+\!\bar{\wb}_k^H\!\Qb_{ki}\bar{\wb}_k\bar{\zeta}_i)\!-\!\sum_{k{\ne}i}\frac{\bar{\wb}_k^H\!\Qb_{ki}\bar{\wb}_k\bar{\zeta}_i}{1\!+\!\bar{\wb}_k^H\!\Qb_{ki}\bar{\wb}_k\bar{\zeta}_i}\notag\\
&~~~~~+\bigg(\sigma_i^2+\sum_{k{\ne}i}\frac{\wb_k^H\Qb_{ki}\wb_k}{1+\bar{\wb}_k^H\Qb_{ki}\bar{\wb}_k\bar{\zeta}_i}\bigg)\cdot\zeta_i\notag\\
&\triangleq\Psi_i(\zeta_i,\{\wb_k\}_{k{\ne}i}\mid\{\bar{\wb}_k\}_{k{\ne}i})\label{eq:xi_lower_bound}
\end{align}

\vspace{-.1cm}\noindent{where}
$\bar{\zeta}_i=\xi_i(\{\bar{\wb}_k\}_{k{\ne}i})$, and the inequality
is due to the first-order approximation of the concave logarithm
function, i.e., $\ln(y)\le\ln(x)+\frac{y-x}{x}$ $\forall x,y\ge0$.
Note that
$\Psi_i(\zeta_i,\{\wb_k\}_{k{\ne}i}\mid\{\bar{\wb}_k\}_{k{\ne}i})$
is a locally tight upper bound of
$\Phi_i(\zeta_i,\{\wb_k\}_{k{\ne}i})$; moreover, similar to
$\Phi_i(\xi_i,\{\wb_k\}_{k{\ne}i})$,
$\Psi_i(\zeta_i,\{\wb_k\}_{k{\ne}i}\mid\{\bar{\wb}_k\}_{k{\ne}i})$
is continuously differentiable w.r.t.
$(\zeta_i,\{\wb_k\}_{k{\ne}i})$, and is strictly increasing w.r.t.
$\zeta_i$. As a result, there exists a unique continuously
differentiable function, denoted by
$\zeta_i(\{\wb_k\}_{k{\ne}i}\mid\{\bar{\wb}_k\}_{k{\ne}i})$, such
that\vspace{-.1cm}
\[
\Psi_i\big(\zeta_i(\{\wb_k\}_{k{\ne}i}|\{\bar{\wb}_k\}_{k{\ne}i}),\{\wb_k\}_{k{\ne}i}\mid\{\bar{\wb}_k\}_{k{\ne}i}\big)=0,
\]

\vspace{-.1cm}\noindent{for} all $\{\wb_k\}_{k{\ne}i}$. In
particular, it follows from \eqref{eq:xi_lower_bound} that
$\zeta_i(\{\wb_k\}_{k{\ne}i}\mid\{\bar{\wb}_k\}_{k{\ne}i})$ has a
closed-form expression as\vspace{-.1cm}
\begin{align}
&\zeta_i(\{\wb_k\}_{k{\ne}i}\mid\{\bar{\wb}_k\}_{k{\ne}i})=\notag\\
&~\gamma_i(\{\bar{\wb}_k\}_{k{\ne}i})\bigg(\!\sigma_i^2\!+\!\sum_{j{\ne}i}\!\frac{\wb_j^H\!\Qb_{ji}\wb_j}{1\!+\!\bar{\wb}_j^H\!\Qb_{ji}\bar{\wb}_j\xi_i(\{\bar{\wb}_k\}_{k{\ne}i})}\!\bigg)^{-1}\label{eq:zeta_CLSFORM}
\end{align}

\vspace{-.2cm}\noindent{where}\vspace{-.1cm}
\begin{align}
\gamma_i(\{\bar{\wb}_k\}_{k{\ne}i})\triangleq&\sum_{j{\ne}i}\frac{\bar{\wb}_j^H\Qb_{ji}\bar{\wb}_j\cdot\xi_i(\{\bar{\wb}_k\}_{k{\ne}i})}{1+\bar{\wb}_j^H\Qb_{ji}\bar{\wb}_j\cdot\xi_i(\{\bar{\wb}_k\}_{k{\ne}i})}-\ln\rho_i\notag\\
&~~-\sum_{j{\ne}i}\ln(1+\bar{\wb}_j^H\Qb_{ji}\bar{\wb}_j\cdot\xi_i(\{\bar{\wb}_k\}_{k{\ne}i}))\notag\\
=&\sum_{j{\ne}i}\frac{\bar{\wb}_j^H\Qb_{ji}\bar{\wb}_j\cdot\xi_i(\{\bar{\wb}_k\}_{k{\ne}i})}{1+\bar{\wb}_j^H\Qb_{ji}\bar{\wb}_j\cdot\xi_i(\{\bar{\wb}_k\}_{k{\ne}i})}\notag\\
&~~+\sigma_i^2\xi_i(\{\bar{\wb}_k\}_{k{\ne}i})~~~~~~~~~~~~~~~~\text{(by
\eqref{eq:xi_def})}\notag\\
>&0,~\forall\{\bar{\wb}_k\}_{k{\ne}i}.\label{eq:gamma_def}
\end{align}
By \eqref{eq:zeta_CLSFORM} and \eqref{eq:gamma_def}, one can see
that $\zeta_i(\{\wb_k\}_{k{\ne}i}\mid\{\bar{\wb}_k\}_{k{\ne}i})>0$
for all feasible $\{\wb_k\}_{k{\ne}i}$ and
$\{\bar{\wb}_k\}_{k{\ne}i}$. Moreover, from
\eqref{eq:xi_lower_bound} and \eqref{eq:zeta_CLSFORM},
$\zeta_i(\{\wb_k\}_{k{\ne}i}\mid\{\bar{\wb}_k\}_{k{\ne}i})$ is a
locally tight lower bound of $\xi_i(\{\wb_k\}_{k{\ne}i})$, i.e.,
\begin{subequations}\label{eq:BSUM_zeta}
\begin{align}
&\zeta_i(\{\wb_k\}_{k{\ne}i}\mid\{\bar{\wb}_k\}_{k{\ne}i})\le\xi_i(\{\wb_k\}_{k{\ne}i}),\label{eq:BSUM_zeta_a}\\
&\zeta_i(\{\bar{\wb}_k\}_{k{\ne}i}\mid\{\bar{\wb}_k\}_{k{\ne}i})=\xi_i(\{\bar{\wb}_k\}_{k{\ne}i}),\label{eq:BSUM_zeta_b}
\end{align}
\end{subequations}
for all $\|\wb_k\|^2\le{P_k}$, $\|\bar{\wb}_k\|^2\le{P_k}$,
$k\ne{i}$. Therefore, the following function
\begin{align}
&\tilde{\Ucal}_{wsr}(\{\wb_k\}\mid\{\bar{\wb}_k\})\triangleq\notag\\
&~~\sum_{i=1}^K\alpha_i\log_2(1+\zeta_i(\{\wb_k\}_{k{\ne}i}|\{\bar{\wb}_k\}_{k{\ne}i})\cdot\wb_i^H\Qb_{ii}\wb_i).\label{def:tildeUwsr}
\end{align}
serves as a locally tight lower bound of the WSR utility
$\Ucal_{wsr}(\{\wb_k\})$ in \eqref{WSRM_a}. By defining
\begin{subequations}\label{def:Qbar}
\begin{align}
&\bar{\Qb}_{ii}(\{\bar{\wb}\}_{k{\ne}i})\!\triangleq\!\gamma_i(\{\bar{\wb}_k\}_{k{\ne}i})\cdot\Qb_{ii},\label{def:Qbar_a}\\
&\bar{\Qb}_{ji}(\{\bar{\wb}\}_{k{\ne}i})\!\triangleq\!\big(1\!+\!\bar{\wb}_j^H\Qb_{ji}\bar{\wb}_j\xi_i(\{\bar{\wb}_k\}_{k{\ne}i})\big)^{-1}\Qb_{ji},\label{def:Qbar_b}
\end{align}
\end{subequations}
for $i,j=1,\dots,K$, $j\ne{i}$, and by \eqref{eq:zeta_CLSFORM}, one
can further express $\tilde{\Ucal}_{wsr}(\cdot\mid\cdot)$ as
\begin{align}
&\tilde{\Ucal}_{wsr}(\{\wb_k\}\mid\{\bar{\wb}_k\})=\notag\\
&~\sum_{i=1}^K\alpha_i\log_2\bigg(1+\frac{\wb_i^H\bar{\Qb}_{ii}\wb_i}{\sigma_i^2+\sum_{j{\ne}i}\wb_j^H\bar{\Qb}_{ji}\wb_j}\bigg),\label{def:tildeUwsr
2}
\end{align}
where $\bar{\Qb}_{ji}(\{\bar{\wb}\}_{k{\ne}i})$ are denoted by
$\bar{\Qb}_{ji}$ for all $i,j=1,\dots,K$, for notational simplicity.
It is interesting to note that
$\tilde{\Ucal}_{wsr}(\{\wb_k\}\mid\{\bar{\wb}_k\})$ in
\eqref{def:tildeUwsr 2} is virtually the WSR of an IFC, and the
ratio in $\log_2(\cdot)$ is the associated
\textit{signal-to-interference-plus-noise-ratio} (SINR). It is known
that SINR is closely related to the minimum mean-square error (MMSE)
of estimated symbols, and this relation has been exploited for
developing efficient precoder optimization algorithms, see, e.g.,
the iterative WMMSE method in \cite{Shi2011_IteMMSE}. Here, we adopt
an idea similar to the one in \cite{Shi2011_IteMMSE} to further
obtain a lower bound of $\tilde{\Ucal}_{wsr}(\cdot\mid\cdot)$ that
is separable over $\{\wb_k\}$.

Consider an $N_t\times{N_t}$ MIMO channel with channel matrix
$\bar{\Qb}_{ii}^{1/2}$, where
$(\bar{\Qb}_{ii}^{1/2})^H\bar{\Qb}_{ii}^{1/2}=\bar{\Qb}_{ii}$, and
additive noise
$\nb_i\sim\mathcal{CN}(\zerob,(\sigma_i^2+\sum_{j\ne{i}}\wb_j^H\bar{\Qb}_{ji}\wb_j)\cdot\mathbf{I}_{N_t})$,
where $\mathbf{I}_{N_t}$ is the $N_t\times{N_t}$ identity matrix.
Suppose that the transmitter sends the information signal $s_i$ to
the receiver via transmit beamforming $\wb_i$, and the receiver
estimates $s_i$ by linear decoder $\yb_i$. Then, the MMSE of the
estimation is given by\vspace{-.1cm}
\begin{align}
&\min_{\yb_i\in\mathbb{C}^{N_t}}\left|1-\yb_i^H\bar{\Qb}_{ii}^{1/2}\wb_i\right|^2+\big(\sigma_i^2+\sum_{j{\ne}i}\wb_j^H\bar{\Qb}_{ji}\wb_j\big)\yb_i^H\yb_i\notag\\
&~~~~=1-\frac{\wb_i^H\bar{\Qb}_{ii}\wb_i}{\sigma_i^2+\sum_{j=1}^K\wb_j^H\bar{\Qb}_{ji}\wb_j}\notag\\
&~~~~=\bigg(1+\frac{\wb_i^H\bar{\Qb}_{ii}\wb_i}{\sigma_i^2+\sum_{j{\ne}i}\wb_j^H\bar{\Qb}_{ji}\wb_j}\bigg)^{-1}.\label{eq:mse_sinr}
\end{align}

\vspace{-.1cm}\noindent{The} optimal $\yb_i$ to \eqref{eq:mse_sinr}
can be shown to be\vspace{-.1cm}
\begin{equation}\label{eq:mse_sinr_opt}
\yb_i=\frac{\bar{\Qb}_{ii}^{1/2}\wb_i}{\sigma_i^2+\sum_{j=1}^K\wb_j^H\bar{\Qb}_{ji}\wb_j}.
\end{equation}

\vspace{-.1cm}\noindent{Then}, we can further obtain a lower bound
of $\tilde{\Ucal}_{wsr}(\{\wb_k\}\mid\{\bar{\wb}_k\})$ as presented
in \eqref{eq:BSUM_lnrz} on the top of the next page,
\begin{figure*}[t]
\begin{align}
&\tilde{\Ucal}_{wsr}(\{\wb_k\}\mid\{\bar{\wb}_k\})\notag\\
&=\sum_{i=1}^K\!-\alpha_i\!\log_2\!\bigg(\!\bigg(\!1\!+\!\frac{\wb_i^H\bar{\Qb}_{ii}\wb_i}{\sigma_i^2\!+\!\sum_{j{\ne}i}\wb_j^H\bar{\Qb}_{ji}\wb_j}\!\bigg)^{-1}\bigg)~~~~~~~~~~~~~~~~~~~~~~~~~~~~~~\text{(by \eqref{def:tildeUwsr 2})}\notag\\
&\ge\sum_{i=1}^K\!-\alpha_i\!\log_2\!\bigg(\!\!\left|1\!-\!\bar{\yb}_i^H\bar{\Qb}_{ii}^{1/2}\wb_i\right|^2\!\!\!+\!\big(\sigma_i^2\!+\!\sum_{j{\ne}i}\wb_j^H\!\bar{\Qb}_{ji}\wb_j\big)\bar{\yb}_i^H\bar{\yb}_i\!\bigg)~~~~~~~~~~~~~\text{(by \eqref{eq:mse_sinr})}\notag\\
&\ge\sum_{i=1}^K\!-\alpha_i\!\log_2\!\bigg(\!\!\left|1\!-\!\bar{\yb}_i^H\bar{\Qb}_{ii}^{1/2}\bar{\wb}_i\right|^2\!\!\!+\!\big(\sigma_i^2\!+\!\sum_{j{\ne}i}\bar{\wb}_j^H\!\bar{\Qb}_{ji}\bar{\wb}_j\big)\bar{\yb}_i^H\bar{\yb}_i\!\bigg)\!+\!\frac{\alpha_i}{\ln2}\!\bigg(\!1\!-\!\frac{|1\!-\!\bar{\yb}_i^H\bar{\Qb}_{ii}^{1/2}\wb_i|^2\!+\!\big(\sigma_i^2\!+\!\sum_{j{\ne}i}\wb_j^H\bar{\Qb}_{ji}\wb_j\big)\bar{\yb}_i^H\bar{\yb}_i}{|1\!-\!\bar{\yb}_i^H\bar{\Qb}_{ii}^{1/2}\bar{\wb}_i|^2\!+\!\big(\sigma_i^2\!+\!\sum_{j{\ne}i}\bar{\wb}_j^H\bar{\Qb}_{ji}\bar{\wb}_j\big)\bar{\yb}_i^H\bar{\yb}_i}\bigg)\notag\\
&\triangleq\bar{\Ucal}_{wsr}(\{\wb_k\}\mid\{\bar{\wb}_k\}),\label{eq:BSUM_lnrz}
\end{align}

\vspace{-.35cm}\hrulefill\vspace{-.3cm}
\end{figure*}
where $\bar{\yb}_i$ is defined as\vspace{-.2cm}
\begin{equation}\label{eq:ybar_def}
\bar{\yb}_i\triangleq\frac{\bar{\Qb}_{ii}^{1/2}\bar{\wb}_i}{\sigma_i^2+\sum_{j=1}^K\bar{\wb}_j^H\bar{\Qb}_{ji}\bar{\wb}_j},~i=1,\dots,K,
\end{equation}
and the second inequality is obtained by the fact that
$-\ln(y)\ge-\ln(x)-\frac{y-x}{x}$ $\forall x,y\ge0$. Moreover, by
\eqref{eq:mse_sinr}, \eqref{eq:mse_sinr_opt} and
\eqref{eq:ybar_def}, one can show that the lower bound
$\bar{\Ucal}_{wsr}(\{\wb_k\}\mid\{\bar{\wb}_k\})$ is actually
locally tight to $\tilde{\Ucal}_{wsr}(\{\wb_k\}\mid\{\bar{\wb}_k\})$
when $\wb_k=\bar \wb_k$ $\forall k=1,\ldots,K,$ i.e.,
$\bar{\Ucal}_{wsr}(\{\bar\wb_k\}\mid\{\bar{\wb}_k\})=\tilde{\Ucal}_{wsr}(\{\bar
\wb_k\}\mid\{\bar{\wb}_k\})$. Combining this result with the fact
that $\tilde{\Ucal}_{wsr}(\{\wb_k\}\mid\{\bar{\wb}_k\})$ is a
locally tight lower bound of the original WSR
$\Ucal_{wsr}(\{\wb_k\})$, we conclude that
$\bar{\Ucal}_{wsr}(\{\wb_k\}\mid\{\bar{\wb}_k\})$ is also a locally
tight lower bound of $\Ucal_{wsr}(\{\wb_k\})$,
satisfying\vspace{-.4cm}
\begin{subequations}\label{eq:regularity_WMMSE_0}
\begin{align}\label{eq:regularity_WMMSE_2}
&\bar{\Ucal}_{wsr}(\{\wb_k\}\mid\{\bar{\wb}_k\})\le\Ucal_{wsr}(\{\wb_k\}),\\
&\bar{\Ucal}_{wsr}(\{\bar{\wb}_k\}\mid\{\bar{\wb}_k\})=\Ucal_{wsr}(\{\bar{\wb}_k\}),\\
&\bar{\Ucal}_{wsr}(\{\wb_k\}\mid\{\bar{\wb}_k\})~\text{is continuous
in}~(\{\wb_k\},\{\bar{\wb}_k\}),
\end{align}
\end{subequations}
for all $\|\wb_k\|^2\le{P_k}$, $\|\bar{\wb}_k\|^2\le{P_k}$,
$k=1,\dots,K$.

Therefore, we can apply the SUM method \cite{Razaviyayn_2013BSUM}
(i.e., BSUM with one block variable $(\wb_1,\ldots,\wb_K)$) to
problem \eqref{WSRM}, by using
$\bar{\Ucal}_{wsr}(\{\wb_k\}\mid\{\bar{\wb}_k\})$ in
\eqref{eq:BSUM_lnrz} as the surrogate function. Specifically,
according to SUM, the beamforming vectors are iteratively updated as
\begin{equation}\label{WSRM_WMMSE_BSUM}
(\wb_1^{[n]},\dots,\wb_K^{[n]})\!=\!\mathrm{arg}\!\!\!\max_{\substack{\|\wb_i\|^2\le{P_i},\\i=1,\dots,K}}\!\bar{\Ucal}_{wsr}(\{\wb_k\}|\{\wb_k^{[n-1]}\})
\end{equation}
By \eqref{eq:regularity_WMMSE_0} and by \cite[Theorem 1]{Razaviyayn_2013BSUM}, the sequence generated by \eqref{WSRM_WMMSE_BSUM} is guaranteed to converge\footnote{For the SUM method, convergence is guaranteed without the need of unique solution to problem \eqref{WSRM_WMMSE_BSUM}; see \cite[Theorem 1]{Razaviyayn_2013BSUM}.}:
\begin{Proposition}\label{prop:WMMSE-BSUM_converge}
Every limit point of
$\{(\wb_1^{[n]},\dots,\wb_K^{[n]})\}_{n=1}^\infty$ generated by
\eqref{WSRM_WMMSE_BSUM} is a stationary point of problem
\eqref{WSRM}.
\end{Proposition}

Unlike the DBSUM algorithm (Algorithm \ref{alg:BSUM_UMX}),
implementation of \eqref{WSRM_WMMSE_BSUM} can be completely parallel
with only a small amount of messages exchanged among the
transmitters. Specifically, because both the surrogate function
$\bar{\Ucal}_{wsr}(\{\wb_k\}|\{\bar{\wb}_k\})$ and the constraint
set are separable over the beamforming vectors $\wb_1,\dots,\wb_K$,
problem \eqref{WSRM_WMMSE_BSUM} can be decomposed into $K$ parallel
subproblems as (see \eqref{eq:BSUM_lnrz})
\begin{align}
\wb_i^{[n]}=\arg~\min_{\|\wb_i\|^2\le{P_i}}&~\eta_i|1-\bar{\yb}_i^H\bar{\Qb}_{ii}^{1/2}\wb_i|^2\notag\\
&~+\sum_{j{\ne}i}\eta_j\big(\wb_i^H\bar{\Qb}_{ij}\wb_i\big)\bar{\yb}_j^H\bar{\yb}_j\label{eq:subproblem}
\end{align}
for $i=1,\ldots,K$, where $\eta_j\triangleq
\frac{\alpha_j}{\ln2}[|1-\bar{\yb}_j^H\bar{\Qb}_{jj}^{1/2}\bar{\wb}_j|^2
+\big(\sigma_j^2+\sum_{k{\ne}j}\bar{\wb}_k^H\bar{\Qb}_{kj}\bar{\wb}_k\big)\bar{\yb}_j^H\bar{\yb}_j]^{-1}$,
$j=1,\ldots,K$. In addition, problem \eqref{eq:subproblem} can be
solved very efficiently, e.g., using the gradient projection method
\cite[Section 2.3.1]{BK:Bertsekas1999} or the Lagrange dual method
\cite{BK:BoydV04}. Finally, we summarize the proposed DWMMSE
algorithm for problem \eqref{WSRM} in Algorithm
\ref{alg:WMMSE_BSUM}.

\begin{algorithm}[t]
\caption{DWMMSE algorithm for problem \eqref{WSRM}}
\begin{algorithmic}[1]\label{alg:WMMSE_BSUM}
\STATE {\bf Input} a set of beamformers $\{\wb_i^{[0]}\}$ satisfying
\eqref{WSRM_b};

\STATE Set $n:=0$;

\REPEAT

\STATE $n:=n+1$;

\STATE Each transmitter $i$ obtains $\wb_i^{[n]}$ by solving
\eqref{eq:subproblem}, and sends
$(\wb_i^{[n]})^H\Qb_{ij}\wb_i^{[n]}$ to transmitter $j$ for all
$j{\ne}i$;

\STATE After receiving the quantities
$(\wb_i^{[n]})^H\Qb_{ij}\wb_i^{[n]}$ $\forall i\neq j$, each
transmitter $j$ sends
$\theta_{ij}=\frac{\eta_j\bar{\yb}_j^H\bar{\yb}_j}{1+\bar{\wb}_i^H\Qb_{ij}\bar{\wb}_i
\cdot\xi_j(\{\bar{\wb}_k\}_{k{\ne}j})}$ to transmitter $i$ for all
$i{\ne}j$, where $\bar{\wb}_k=\wb_k^{[n]}$, $\forall{k}$, and
$\theta_{ij}\Qb_{ij}=\eta_j\bar{\Qb}_{ij}\bar{\yb}_j^H\bar{\yb}_j$
(cf. \eqref{def:Qbar_b}), $\forall{i,j}$, $i\ne{j}$;

\UNTIL the predefined stopping criterion is met.

\STATE {\bf Output} $\{\wb_i^{[n]}\}$ as an approximate solution of
\eqref{WSRM}.
\end{algorithmic}
\end{algorithm}

\ \\

\vspace{-1.5cm}
\vspace{-.2cm}\section{Outer Approximation by Polyblock
Optimization}\label{sec:polyblock_outer_apprx}

The DBSUM algorithm and the DWMMSE algorithm (that are based on BSUM
and SUM methods \cite{Razaviyayn_2013BSUM}, respectively) presented
in the previous two sections are so called ``inner'' approximation
methods \cite{Marks1978} since, at each iteration, the approximate
beamforming solutions are restrictively feasible and provide lower
bounds to problem \eqref{UMX_CDI_Pr}. In this section, we consider
an ``outer'' approximation method that instead solves an
constraint-relaxed version of problem \eqref{UMX_CDI_Pr}, thus
providing upper bounds to the optimal value of problem
\eqref{UMX_CDI_Pr}. The motive is that the proposed DBSUM and WMMSE
algorithms can be benchmarked against such a method, as the
approximation errors of the proposed algorithms are no larger than
the gap between the outer and inner approximation methods. Compared
with the exhaustive search method which is not feasible when the
number of users is large, the outer approximation method is
computationally more efficient.

Our approach is based on the polyblock outer approximation (POA)
algorithm
\cite{Tuy2000_Monotonic,Jorswieck2010_POA,Utschick2012_POA,Bjornson2012_BRB,Zhang_2012_POA},
which is used for solving the \textit{monotonic optimization
problems}\cite{Tuy2000_Monotonic}. To be self-contained, a review of
the POA algorithm is given in Appendix \ref{sec:intr_MonoOpt_POA}.
Roughly speaking, the POA algorithm systematically constructs a
sequence of optimization problems which has a structured feasible
set (called polyblock; see Definition \ref{def:polyblock} in
Appendix \ref{sec:intr_MonoOpt_POA}) that contains the feasible set
of the original problem. The structured feasible set shrinks at
every iteration and converges to the true feasible set of the
original problem. Thereby, the objective values of the constructed
problems converge to the true optimal value from above
asymptotically.

Recall the outage-constrained problem \eqref{UMX_CDI_Pr}. By
\eqref{eq:outage_constraint_CLSFORM} and \eqref{eq:xi_def}, problem
\eqref{UMX_CDI_Pr} can be compactly written as\vspace{-.1cm}
\begin{subequations}\label{UMX_mono_opt}
\begin{align}
\max_{\substack{R_i\ge0,\\i=1,\dots,K}}~&U(R_1,\dots,R_K)\\
\text{s.t.}~&[R_1,\dots,R_K]^T\in\mathcal{R}\triangleq
\bigcup_{\substack{\|\wb_i\|^2{\le}P_i,\\i=1,\dots,K}}\mathcal{R}(\{\wb_k\}),
\end{align}
\end{subequations}

\vspace{-.3cm}\noindent{where}
\begin{align}
&\mathcal{R}(\{\wb_k\})\triangleq\notag\\
&~\bigg\{\![R_1,\dots,R_K]^T\!\!\succeq\!\zerob\left|\Phi_i\!\left(\!\frac{2^{R_i}-1}{\wb_i^H\Qb_{ii}\wb_i},\{\wb_k\}_{k{\ne}i}\right)\!\le\!0,~\forall{i}\!\right.\bigg\}.\label{achievable_rate_region}
\end{align}

\vspace{-.2cm}\noindent{By} the fact that
$\Phi_i\left(\frac{2^{R_i}-1}{\wb_i^H\Qb_{ii}\wb_i},\{\wb_k\}_{k{\ne}i}\right)$
is increasing w.r.t. $R_i$, one can easily verify that
$\mathcal{R}(\{\wb_k\})$ is a normal set; thus
$\mathcal{R}\subseteq\mathbb{R}_+^K$, which is the union of normal
sets, is also a normal set \cite[Proposition 3]{Tuy2000_Monotonic}.
As a result, problem \eqref{UMX_mono_opt} is a monotonic
optimization problem. However, directly applying the POA algorithm
(Algorithm \ref{alg:poa} in Appendix \ref{sec:intr_MonoOpt_POA}) to
problem \eqref{UMX_mono_opt} results in prohibitively high
computational complexity. In particular, both step 3 and step 7 of
Algorithm \ref{alg:poa} for problem \eqref{UMX_mono_opt} corresponds
to solving a problem of the form\vspace{-.1cm}
\begin{align}
&\,~~~~\max_{\beta\ge0}~U(\beta{v_1^\star},\dots,\beta{v_K^\star})\notag\\
&\,~~~~~~~\text{s.t.}~[\beta{v_1^\star},\dots,\beta{v_K^\star}]^T\in\mathcal{R}.\notag\\
=~&\max_{\substack{\beta\ge0,\wb_i\in\mathbb{C}^{N_t},\\i=1,\dots,K}}~\beta \label{UMX_mono_opt_intsct}\\
&\,~~~~~~~\text{s.t.}~\Phi_i\bigg(\frac{2^{\beta{v_i^\star}}-1}{\wb_i^H\Qb_{ii}\wb_i},\{\wb_k\}_{k{\ne}i}\bigg)\le0,\notag\\
&\,~~~~~~~~~~~\|\wb_i\|^2{\le}P_i,~i=1,\dots,K,\notag
\end{align}

\vspace{-.05cm}\noindent{where}
$\vb^\star=[v_1^\star,\dots,v_K^\star]^T\succeq\zerob$ is a given
point, and the equality is due to the fact that the utility
$U(\cdot)$ is nondecreasing. As seen, problem
\eqref{UMX_mono_opt_intsct} is equivalent to problem
\eqref{UMX_CDI_Pr} with the MMF rate utility, which, however, is
NP-hard in general (when $N_t\geq 2$) as proved in \cite[Theorem
3]{Li_14_cplx}. Hence, it is inefficient to use the POA algorithm to
solve problem \eqref{UMX_mono_opt}.

To overcome this issue, we instead consider a relaxed convex
approximation problem. Let us consider a lower bound of
$\Phi_i\big(\frac{2^{R_i}-1}{\wb_i^H\Qb_{ii}\wb_i},\{\wb_k\}_{k{\ne}i}\big)$
(cf. \eqref{eq:xi_def}) as follows\vspace{-.05cm}
\begin{align}
&\Phi_i\bigg(\frac{2^{R_i}-1}{\wb_i^H\Qb_{ii}\wb_i},\{\wb_k\}_{k{\ne}i}\bigg)\notag\\
&~~\ge\ln\rho_i+\ln\bigg(1+\frac{(2^{R_i}-1)\sigma_i^2}{\wb_i^H\Qb_{ii}\wb_i}\bigg)\notag\\
&~~~~~~~~~~~+\sum_{k\ne{i}}\ln\bigg(1+\frac{(2^{R_i}-1)\wb_k^H\Qb_{ki}\wb_k}{\wb_i^H\Qb_{ii}\wb_i}\bigg)\notag\\
&~~=\ln\bigg[\rho_i\times\bigg(1+\frac{(2^{R_i}-1)\sigma_i^2}{\wb_i^H\Qb_{ii}\wb_i}\bigg)\notag\\
&~~~~~~~~~~~\times\prod_{k\ne{i}}\bigg(1+\frac{(2^{R_i}-1)\wb_k^H\Qb_{ki}\wb_k}{\wb_i^H\Qb_{ii}\wb_i}\bigg)\bigg],\label{eq:relax_1}
\end{align}
for $i=1,\dots,K$, where the inequality is owing to $x\ge\ln(1+x)$
$\forall x\geq 0$. Moreover, since the terms
$\frac{(2^{R_i}-1)\sigma_i^2}{\wb_i^H\Qb_{ii}\wb_i}$ and
$\frac{(2^{R_i}-1)\wb_k^H\Qb_{ki}\wb_k}{\wb_i^H\Qb_{ii}\wb_i}$
$\forall{k\ne{i}}$ are non-negative, we can further obtain
\begin{align}
&\ln\bigg[\rho_i\cdot\bigg(1\!+\!\frac{(2^{R_i}\!-\!1)\sigma_i^2}{\wb_i^H\Qb_{ii}\wb_i}\bigg)\!\prod_{k\ne{i}}\!\bigg(\!1\!+\!\frac{(2^{R_i}\!-\!1)\wb_k^H\Qb_{ki}\wb_k}{\wb_i^H\Qb_{ii}\wb_i}\!\bigg)\!\bigg]\notag\\
&~\ge\ln\bigg[\rho_i\cdot\bigg(\!1\!+\!\frac{(2^{R_i}\!-\!1)\sigma_i^2}{\wb_i^H\Qb_{ii}\wb_i}+\sum_{k\ne{i}}\frac{(2^{R_i}\!-\!1)\wb_k^H\Qb_{ki}\wb_k}{\wb_i^H\Qb_{ii}\wb_i}\!\bigg)\!\bigg]\notag\\
&~=\ln\bigg[\rho_i\cdot\bigg(1+\frac{\sigma_i^2+\sum_{k\ne{i}}\tr(\wb_k^H\Qb_{ki}\wb_k)}{(2^{R_i}-1)^{-1}\tr(\wb_i^H\Qb_{ii}\wb_i)}\bigg)\bigg]\label{eq:relax_2}
\end{align}
for $i=1,\dots,K$. By using the lower bound in \eqref{eq:relax_2},
we obtain the following problem which has a relaxed constraint set
comparing to problem \eqref{UMX_CDI_Pr}\vspace{-.1cm}
\begin{subequations}\label{UMX_mono_opt_rlx2}
\begin{align}
&\max_{\substack{\wb_i\in\mathbb{C}^{N_t}\!,~R_i\ge0,\\i=1,\dots,K}}~U(R_1,\dots,R_K)\label{UMX_mono_opt_rlx2_a}\\
&~~~~~~~~~~\text{s.t.}~\frac{\sigma_i^2\!+\!\sum_{k{\ne}i}\!\tr(\wb_k\wb_k^H\!\Qb_{ki})}{(2^{R_i}-1)^{-1}\tr(\wb_i\wb_i^H\Qb_{ii})}\le\frac{1\!-\!\rho_i}{\rho_i},\label{UMX_mono_opt_rlx2_b}\\
&~~~~~~~~~~~~~~~\tr(\wb_i\wb_i^H){\le}P_i,~i=1,\dots,K.\label{UMX_mono_opt_rlx2_c}
\end{align}
\end{subequations}

\vspace{-.1cm}\noindent{Furthermore}, we consider the
\textit{semidefinite relaxation} (SDR) technique \cite{Luo2010_SPM},
by which we relax the rank-one $\wb_i\wb_i^H$ to a PSD matrix
$\Wb_i\succeq\zerob$, for all $i=1,\ldots,K$. The resultant problem
can be expressed as\vspace{-.1cm}
\begin{subequations}\label{UMX_mono_opt_rlx}
\begin{align}
\max_{\substack{R_i\ge0,\\i=1,\dots,K}}~&U(R_1,\dots,R_K)\\
\text{s.t.}~&[R_1,\dots,R_K]^T\!\in\!\tilde{\mathcal{R}}\!\triangleq\!\!\bigcup_{\substack{\tr(\Wb_i){\le}P_i,\\\Wb_i\succeq\zerob,~\forall{i}}}\!\!\tilde{\mathcal{R}}(\{\Wb_k\}),
\end{align}
\end{subequations}

\vspace{-.3cm}\noindent{where}\vspace{-.1cm}
\begin{align}
&\tilde{\mathcal{R}}(\{\Wb_k\})\triangleq\notag\\
&\left\{\![R_1,\dots,R_K]^T\!\succeq\!\zerob\left|\frac{\sigma_i^2\!+\!\sum_{k{\ne}i}\!\tr(\Wb_k\Qb_{ki})}{(2^{R_i}\!-\!1)^{-1}\tr(\Wb_i\Qb_{ii})}\!\le\!\frac{1-\rho_i}{\rho_i},\forall{i}\right.\right\}.\label{achievable_rate_region_rlx}
\end{align}

\vspace{-.1cm}\noindent{Note} that
$\mathcal{R}\subseteq\tilde{\mathcal{R}}\subseteq\mathbb{R}_+^K$,
i,e., problem \eqref{UMX_mono_opt_rlx} is a relaxed problem of
problem \eqref{UMX_mono_opt}. Problem \eqref{UMX_mono_opt_rlx} is a
monotonic optimization problem as $\tilde{\mathcal{R}}$ can be
verified to be normal. Moreover, compared to \eqref{UMX_mono_opt},
problem \eqref{UMX_mono_opt_rlx} can be handled by the POA algorithm
in a more efficient manner. Specifically, step 3 and step 7 of
Algorithm \ref{alg:poa}, which is in Appendix
\ref{sec:intr_MonoOpt_POA}, for problem \eqref{UMX_mono_opt_rlx} now
correspond to solving\vspace{-.2cm}
\begin{align}
&\max_{\substack{\beta\ge0,\Wb_i\in\mathbb{C}^{N_t \times N_t},\\i=1,\dots,K}}~\beta \label{UMX_mono_opt_rlx_intsct}\\
&~~~~~~~~~~~\text{s.t.}~\frac{\sigma_i^2+\sum_{k{\ne}i}\tr(\Wb_k\Qb_{ki})}{(2^{\beta v_i^\star}-1)^{-1}\tr(\Wb_i\Qb_{ii})}\le\frac{1-\rho_i}{\rho_i},\notag\\
&~~~~~~~~~~~~~~~\tr(\Wb_i){\le}P_i,~\Wb_i\succeq\zerob,i=1,\dots,K,\notag
\end{align}

\vspace{-.1cm}\noindent{which} can be shown efficiently solvable by
a bisection method \cite{BK:BoydV04}. In Algorithm
\ref{alg:poa_UMX}, we summarize the POA algorithm for solving
problem \eqref{UMX_mono_opt_rlx} to obtain an upper bound of the
optimal utility value of problem \eqref{UMX_CDI_Pr}.

\vspace{-.1cm}
\section{Simulation Results}\label{sec:simulation}

In this section, we evaluate the performance of Algorithm
\ref{alg:BSUM_UMX} and Algorithm \ref{alg:WMMSE_BSUM} by
simulations. The noise powers at all receivers are assumed to be the
same, i.e., $\sigma_1^2=\cdots=\sigma_K^2\triangleq\sigma^2$, and
all the power constraints are set to one, i.e., $P_1=\cdots=P_K=1$.
The channel covariance matrices $\{\Qb_{ik}\}$ are randomly
generated with full column rank, and with the maximal eigenvalues of
$\{\Qb_{ik}\}$ normalized to $\lambda_{\max}(\Qb_{ii})=1$,
$\lambda_{\max}(\Qb_{ik})=\eta$ for all $k\ne{i}$, $i=1,\dots,K$.
The parameter $\eta\in(0,1]$, thereby, represents the relative
cross-link interference level. The tolerable outage probabilities
are set to $10\%$ for all receivers, i.e.,
$\epsilon_1=\cdots=\epsilon_K=0.1$. The stopping conditions of
Algorithm \ref{alg:BSUM_UMX} and Algorithm \ref{alg:WMMSE_BSUM} are
\begin{subequations}\label{stopping_criterion}
\begin{align}
&\left|\Ucal(\{\wb_k^{[n]}\})\!-\!\Ucal(\{\wb_k^{[n-K]}\})\right|\!<\!10^{-3}\left|\Ucal(\{\wb_k^{[n-K]}\})\right|;\\
&\left|\Ucal(\{\wb_k^{[n]}\})\!-\!\Ucal(\{\wb_k^{[n-1]}\})\right|\!<\!10^{-3}\left|\Ucal(\{\wb_k^{[n-1]}\})\right|,
\end{align}
\end{subequations}
respectively. Note that the DSCA and SCA algorithms in \cite{Li_13}
are also subject to the same stopping conditions as in
\eqref{stopping_criterion}, respectively. The four algorithms
(DBSUM, DWMMSE, DSCA and SCA) are all initialized by randomly
generated unit-norm complex vectors, i.e., $\|\wb_i^{[0]}\|=1$, for
all $i=1,\ldots,K$. Besides, we also run the POA algorithm
(Algorithm \ref{alg:poa_UMX}) as it can yield an upper bound to
problem \eqref{UMX_CDI_Pr}. The subproblem involved in step 3 and
the one in step 7 are handled by the convex solver \texttt{CVX}
\cite{cvx}, and Algorithm \ref{alg:poa_UMX} is stopped if it either
has spent $200$ iterations or has reached the solution accuracy of
$\delta=10^{-3}$. All simulation results are averaged over 500
realizations of CDI $\{\Qb_{ik}\}$.

\begin{algorithm}[t]
\caption{POA algorithm for solving problem \eqref{UMX_mono_opt_rlx}}
\begin{algorithmic}[1]\label{alg:poa_UMX}
\STATE {\bf Initialization:} Set the solution accuracy as
$\delta\ge0$, and set $n:=0$.

\STATE Set
$\mathcal{V}[0]:=\vb^\star[0]\triangleq[v_1^\star[0],\dots,v_K^\star[0]]^T$,
where
$v_i^\star[0]=\log_2(1+\ln(1/\rho_i)P_i\lambda_{\max}(\Qb_{ii})/\sigma_i^2)$,
is the maximal achievable rate of user $i$, for $i=1,\dots,K$;

\STATE Solve problem \eqref{UMX_mono_opt_rlx_intsct} with
$\vb^\star=\vb^\star[0]$ by bisection to obtain $\beta^\star[0]$,
and set $\tilde{\vb}[0]=\beta^\star[0]\vb^\star[0]$;

\WHILE{$U(v_1^\star[n],\dots,v_K^\star[n])-U(\tilde{v}_1[n],\dots,\tilde{v}_K[n])>\delta$}

\STATE $n:=n+1$;

\STATE Set
$\mathcal{V}[n]=\{\mathcal{V}[n-1]\backslash\{\vb^\star[n-1]\}\}\bigcup\{\vb^\star[n-1]-(v_i^\star[n-1]-\tilde{v}_i[n-1])\eb_i\}_{i=1}^K$,
where $\eb_i$ is the $i$th column of the $K{\times}K$ identity
matrix;

\STATE Find
$\vb^\star[n]=\mathrm{arg}{\displaystyle\max_{\vb\in\mathcal{V}[n]}}U(v_1,\dots,v_K)$
followed by solving problem \eqref{UMX_mono_opt_rlx_intsct} with
$\vb^\star=\vb^\star[n]$
by bisection to obtain 
$\tilde{\vb}[n]=\beta^\star[n]\vb^\star[n]$;

\ENDWHILE

\STATE {\bf Output}
  $U(v_1^\star[n],\dots,v_K^\star[n])$ as the approximation
  of the optimal value of \eqref{UMX_mono_opt_rlx}.
\end{algorithmic}
\end{algorithm}

\begin{figure}[t]
\begin{center}
\subfigure[][]{\resizebox{.45\textwidth}{!}{\includegraphics{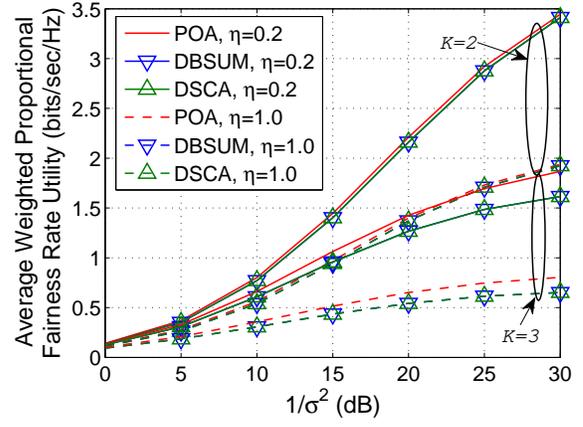}}}
\subfigure[][]{\resizebox{.45\textwidth}{!}{\includegraphics{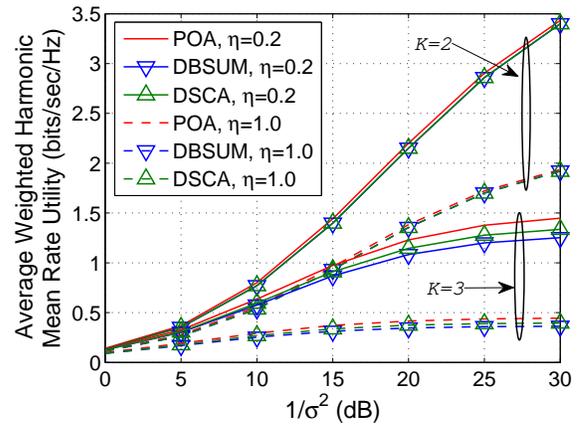}}}
\end{center}\vspace{-0.4cm}
\caption{Performance comparison for the proposed DBSUM algorithm
(Algorithm \ref{alg:BSUM_UMX}) and the DSCA algorithm, for $K=2$,
$(\alpha_1,\alpha_2)=(\frac{1}{2},\frac{1}{2})$, $N_t=4$, and for
$K=3$,
$(\alpha_1,\alpha_2,\alpha_3)=(\frac{1}{6},\frac{1}{3},\frac{1}{2})$,
$N_t=4$; (a) average proportional fairness utility and (b) average
harmonic mean utility versus
$1/\sigma^2$.}\label{fig:Utlt_SNR}\vspace{-0.6cm}
\end{figure}

{\bf Example 1:} We demonstrate the efficacy of Algorithm
\ref{alg:BSUM_UMX}, i.e., the DBSUM algorithm, by comparing it with
the DSCA algorithm in \cite{Li_13} and the benchmark POA algorithm.
We first consider the cases of $K=2$ and $K=3$, and the number of
transmit antennas is set to $N_t=4$. The priority weights are set as
$(\alpha_1,\alpha_2)=(\frac{1}{2},\frac{1}{2})$ and
$(\alpha_1,\alpha_2,\alpha_3)=(\frac{1}{6},\frac{1}{3},\frac{1}{2})$
for the $K=2$ and $K=3$ cases, respectively. Figure
\ref{fig:Utlt_SNR}(a) shows some simulation results for the weighted
proportional fairness rate utility. One can observe from Figure
\ref{fig:Utlt_SNR}(a) that the DBSUM algorithm and the DSCA
algorithm almost yield the same proportional fairness rate for both
$K=2$ and $K=3$, and for both $\eta=0.2$ (the weak interference
scenario) and $\eta=1.0$ (the strong interference scenario). It can
also be observed that, for the case of $K=2$, the DBSUM algorithm
and the DSCA algorithm almost achieve the performance upper bound
obtained by the POA algorithm, implying that both of them can
achieve near optimal performance. For the case of $K=3$, a
non-negligible performance gap between the POA upper bound and the
DBSUM and DSCA algorithms can be observed\footnote{We found in
simulations that under this setting the POA algorithm in general
cannot reach the preset solution accuracy within 200 iterations. So
the performance gap might be reduced if one allows more iterations
for the POA algorithm.}. Nevertheless, both the DBSUM and the DSCA
algorithms can achieve at least $80\%$ of the upper bound,
indicating that the performance loss must be within $20\%$ compared
with the global optimum to problem \eqref{UMX_CDI_Pr}.

Figure \ref{fig:Utlt_SNR}(b) displays some simulation results for
the weighted harmonic mean rate utility. One can observe that, for
the case of $K=2$, the DBSUM and the DSCA almost achieve the optimal
performance; while for the case of $K=3$, the DSCA algorithm
performs slightly better than the DBSUM algorithm, though both
algorithms achieve at least $85\%$ of the optimal harmonic mean
rate.

\begin{figure}[t]
\begin{center}
\subfigure[][]{\resizebox{.45\textwidth}{!}{\includegraphics{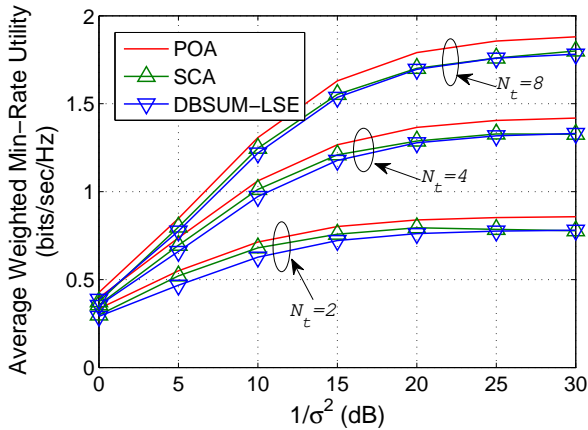}}}
\subfigure[][]{\resizebox{.45\textwidth}{!}{\includegraphics{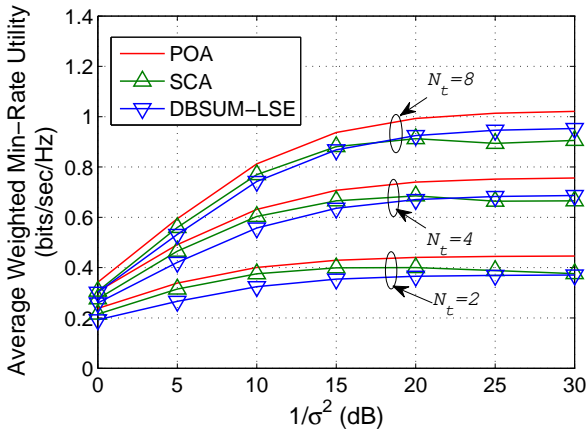}}}
\end{center}\vspace{-0.4cm}
\caption{Simulation results of average achievable weighted min-rate
utility versus $1/\sigma^2$, for (a) $\eta=0.5$, (b) $\eta=1.0$,
where $K=4$, $\alpha_1=\cdots=\alpha_4=\frac{1}{4}$,
$N_t=2,4,8$.}\label{fig:MR_SNR}\vspace{-0.6cm}
\end{figure}

{\bf Example 2:} In Figure \ref{fig:MR_SNR}, we demonstrate the
efficacy of the DBSUM algorithm for handling the MMF rate utility.
Since the log-sum-exp approximation is used, we denote it by
DBSUM-LSE in Figure \ref{fig:MR_SNR}. We consider a $4$-user MISO
IFC under a medium interference level $\eta=0.5$ [Figure
\ref{fig:MR_SNR}(a)] and a strong interference level $\eta=1.0$
[Figure \ref{fig:MR_SNR}(b)], respectively. The user priority
weights are set to be $\alpha_1=\cdots=\alpha_4=\frac{1}{4}$, and
$\gamma=5$ is used in the log-sum-exp approximation (see
\eqref{eq:log_sum_exp_apprx}). Note that the DSCA algorithm is not
able to handle the MMF rate function, so we instead compare
DBSUM-LSE with the centralized SCA algorithm \cite{Li_13}. It is
also worthwhile to note that, for the MMF formulation, the POA
algorithm reduces to solving problem \eqref{UMX_mono_opt_rlx_intsct}
only once, with
$[v_1^\star,\dots,v_K^\star]^T=[\alpha_1,\dots,\alpha_K]^T$. From
both Figure \ref{fig:MR_SNR}(a) and Figure \ref{fig:MR_SNR}(b), one
can see that the SCA algorithm performs slightly better than the
DBSUM-LSE algorithm at low SNR, whereas the two algorithms perform
comparably at high SNR. By comparing with the POA algorithm, both
DBSUM-LSE and SCA algorithms achieve at least $80\%$ of the optimal
MMF rate. It can also be observed that the achievable MMF rate
saturates at high SNR due to the strict user fairness requirement;
however, it can be improved as the number of transmit antennas
increases.

\begin{figure}[t]
\begin{center}
\subfigure[][]{\resizebox{.45\textwidth}{!}{\includegraphics{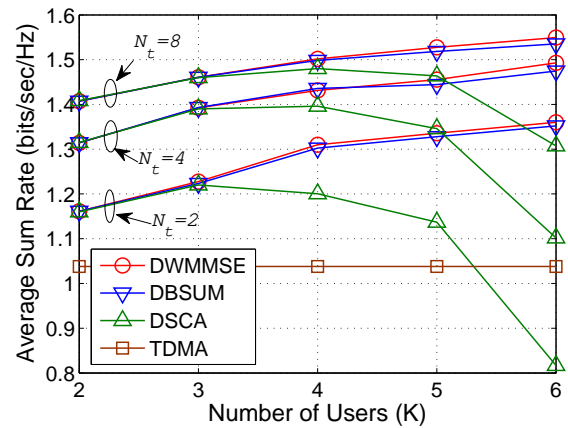}}}
\subfigure[][]{\resizebox{.45\textwidth}{!}{\includegraphics{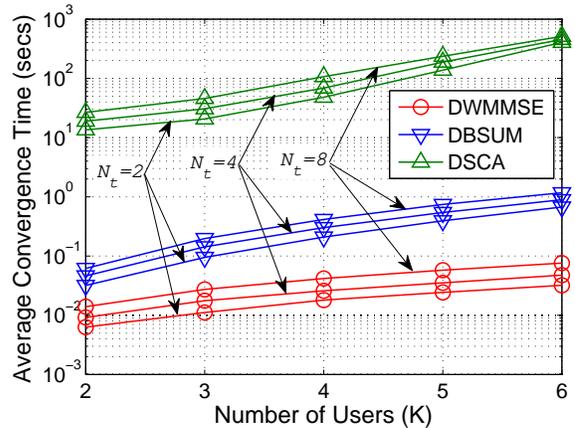}}}
\end{center}\vspace{-0.4cm}
\caption{Performance and complexity comparison for DWMMSE, DBSUM,
and DSCA algorithms, where $1/\sigma^2=10$ dB, $\eta=0.5$,
$\alpha_1=\cdots=\alpha_K=1$, $\rank(\Qb_{ki})=N_t$ for all $k,i$.
(a) Average sum rate versus number of users ($K$), and (b) average
time consumption versus number of users
($K$).}\label{fig:SR_K}\vspace{-0.6cm}
\end{figure}

{\bf Example 3:} In this example, we consider the sum rate utility,
and compare the performance and complexity of the DBSUM algorithm,
the DWMMSE algorithm (Algorithm \ref{alg:WMMSE_BSUM}) and the DSCA
algorithm. To demonstrate the scalability of the DBSUM algorithm and
the DWMMSE algorithm, we consider scenarios for multiple users
($K=2,3,\dots,6$) and multiple transmit antennas ($N_t=2,4,8$). The
SNR and relative cross-link interference level are respectively
fixed to $1/\sigma^2=10$ dB and $\eta=0.5$.

In Figure \ref{fig:SR_K}(a), it can be observed that the DBSUM
algorithm and the DWMMSE algorithm yield nearly the same system
throughput, which increases with the number of users and the number
of transmit antennas. However, the performance of the DSCA algorithm
drastically degrades when $K\ge4$. The reason for this might be that
the DSCA algorithm are relatively easier to get trapped in some
local maximum when $K\ge4$. In Figure \ref{fig:SR_K}(a), the curve
denoted by TDMA represents the achieved system throughput by
time-division multiple access. One can see from this figure that
allowing all the users to access the spectrum simultaneously leads
to higher spectral efficiency than TDMA even when only CDI is
available at the transmitters. As also observed, the performance
gain of the spectrum sharing policy over the TDMA policy increases
with the number of users.

In Figure \ref{fig:SR_K}(b), we compare the computation load of the
three algorithms under test in terms of the average computation time
per realization (in seconds). In our simulations, the convex
subproblems involved in the DSCA algorithm (i.e., \cite[Eqn.
(36)]{Li_13}) and the DBSUM algorithm (i.e., \eqref{BSUM}) are
handled by \texttt{CVX} and the gradient projection method,
respectively; while the subproblem \eqref{eq:subproblem} in the
DWMMSE algorithm is solved by the Lagrange dual method
\cite{BK:BoydV04} (see \cite[Problem (14)]{Shi2011_IteMMSE} for the
details). It can be observed that the average computation time of
the DBSUM algorithm and the DWMMSE algorithm increase at a slower
rate than that of the DSCA algorithm w.r.t. the number of users,
demonstrating that the DBSUM and DWMMSE algorithms have better
scalability. Apart from that, we see from Figure \ref{fig:SR_K}(b)
that the DBSUM algorithm is $10^2\sim10^3$ faster than the DSCA
algorithm, and the DWMMSE algorithm is about ten times faster than
the DBSUM algorithm\footnote{Since the DWMMSE algorithm can only be
implemented sequentially in the computer, the actual computation
time of the DWMMSE algorithm in a parallel system would be even
shorter.}.

\section{Conclusions}\label{sec:conclusion}

We have presented two efficient distributed algorithms for handling
the NP-hard rate outage constrained CoBF design problem in
\eqref{UMX_CDI_Pr}, namely, the DBSUM algorithm (Algorithm
\ref{alg:BSUM_UMX}) and the DWMMSE algorithm (Algorithm
\ref{alg:WMMSE_BSUM}). The former is a Gauss-Seidel type algorithm,
which can handle problem \eqref{UMX_CDI_Pr} with general utility
functions, while the latter is a Jocobi-type algorithm specifically
designed for the weighted sum rate maximization. For the performance
evaluation of the proposed two algorithms, we have also presented a
POA algorithm (Algorithm \ref{alg:poa_UMX}) to obtain an upper bound
to the optimal utility value of problem \eqref{UMX_CDI_Pr}. The
presented simulation results have shown that the proposed DBSUM and
DWMMSE algorithms outperform the existing DSCA algorithm in both
efficacy and computational efficiency, and yield promising
approximation performance as the performance gap from the benchmark
POA algorithm is small (less than $20\%$).

\appendices {\setcounter{equation}{0}
\renewcommand{\theequation}{A.\arabic{equation}}

\section{Proof of Lemma \ref{lemma:rate_func_convexity}}\label{sec:proof_convexity}

For ease of exposition, let us define
$I_{ki}\triangleq\wb_i^H\Qb_{ki}\wb_i$ for $i,k=1,\dots,K$, and set
\begin{align*}
&\tilde{\Phi}_i(\tilde{\xi}_i,\{I_{ki}\}_{k\ne{i}})=\Phi(\tilde{\xi}_i,\{\wb_k\}_{k\ne{i}}),~~~~~~~~\text{(see \eqref{eq:xi_def})}\\
&\tilde{\xi}_i(\{I_{ki}\}_{k{\ne}i})=\xi_i(\{\wb_k\}_{k{\ne}i}),\\
&\tilde{R}_i(\{I_{ki}\}_k)=R_i(\{\wb_k\}),
\end{align*}
for all $i,k=1,\dots,K$. Hence, our goal is to show that
$\tilde{R}_i(\{I_{kI}\}_k)$ is strictly increasing and strictly
concave w.r.t. $I_{ii}$ while is nonincreasing and convex w.r.t.
$I_{ki}$, $k{\ne}i$, for each $i=1,\dots,K$.

Since
$\tilde{\xi}_i(\{I_{ki}\}_{k{\ne}i})=\xi_i(\{\wb_k\}_{k{\ne}i})>0$
for any $\{\wb_k\}_{k{\ne}i}$, it can be directly inferred from the
strict monotoninicity and strict concavity of $\log_2(\cdot)$ that
$\tilde{R}_i(\{I_{ki}\}_k)$ is strictly increasing and strictly
concave w.r.t. $I_{ii}$. To prove the monotonicity and convexity of
$\tilde{R}_i(\{I_{ki}\}_{k{\ne}i})$ w.r.t. $I_{ki}$, $k{\ne}i$, we
need the following lemma:

\begin{Lemma}\label{lemma:implicit_func_monotonicity}
For all $I_{\ell{i}}\ge0$, $\ell{\ne}i$,
$\tilde{\xi}_i(\{I_{ki}\}_{k{\ne}i})$ is strictly decreasing while
$I_{\ell{i}}\cdot\tilde{\xi}_i(\{I_{ki}\}_{k{\ne}i})$ is strictly
increasing w.r.t. $I_{\ell{i}}$, for $i=1,\dots,K$.
\end{Lemma}

\emph{Proof:} By definition, we know that
\begin{align*}
&\tilde{\Phi}_i(\tilde{\xi}_i(\{I_{ki}\}_{k{\ne}i}),\{I_{ki}\}_{k{\ne}i})=\\
&~~\ln\rho_i\!+\!\sigma_i^2\tilde{\xi}_i(\{I_{ki}\}_{k{\ne}i})\!+\!\sum_{k\ne{i}}\ln(1+I_{ki}\tilde{\xi}_i(\{I_{ki}\}_{k{\ne}i}))\!=\!0,
\end{align*}
for all $I_{ki}\ge0$, $k{\ne}i$. Suppose that
$I_{\ell{i}}<I_{\ell{i}}'$. Then,
\begin{align*}
0&=\ln\rho_i+\tilde{\xi}_i'\sigma_i^2+\ln(1+\tilde{\xi}_i'I_{\ell{i}}')+\sum_{k\ne{i},k\ne\ell}\ln(1+\tilde{\xi}_i'I_{ki})\\
&=\ln\rho_i+\tilde{\xi}_i\sigma_i^2+\ln(1+\tilde{\xi}_iI_{\ell{i}})+\sum_{k\ne{i},k\ne\ell}\ln(1+\tilde{\xi}_iI_{ki})\\
&<\ln\rho_i+\tilde{\xi}_i\sigma_i^2+\ln(1+\tilde{\xi}_iI_{\ell{i}}')+\sum_{k\ne{i},k\ne\ell}\ln(1+\tilde{\xi}_iI_{ki}),
\end{align*}
where we denote $\tilde{\xi}_i(\{I_{ki}\}_{k{\ne}i})$ and
$\xi_i(\{I_{ki}\}_{k\ne{i},k\ne\ell},I_{\ell{i}}')$ by
$\tilde{\xi}_i$ and $\tilde{\xi}_i'$ for notational simplicity.
Since $\tilde{\Phi}_i(\tilde{\xi}_i,\{I_{ki}\}_{k{\ne}i})$ is a
strictly increasing function of $\tilde{\xi}_i$, the above
inequality implies $\tilde{\xi}_i>\tilde{\xi}_i'$. Hence,
$\tilde{\xi}_i(\{I_{ki}\}_{k{\ne}i})$ is strictly decreasing w.r.t.
$I_{\ell{i}}$ for all $\ell\ne{i}$. Furthermore, by the fact that
$\tilde{\xi}_i>\tilde{\xi}_i'$, we can obtain
\begin{align*}
0&=\ln\rho_i+\tilde{\xi}_i\sigma_i^2+\ln(1+\tilde{\xi}_iI_{\ell{i}})+\sum_{k\ne{i},k\ne\ell}\ln(1+\tilde{\xi}_iI_{ki})\\
&=\ln\rho_i+\tilde{\xi}_i'\sigma_i^2+\ln(1+\tilde{\xi}_i'I_{\ell{i}}')+\sum_{k\ne{i},k\ne\ell}\ln(1+\tilde{\xi}_i'I_{ki})\\
&<\ln\rho_i+\tilde{\xi}_i\sigma_i^2+\ln(1+\tilde{\xi}_i'I_{\ell{i}}')+\sum_{k\ne{i},k\ne\ell}\ln(1+\tilde{\xi}_iI_{ki}),
\end{align*}
which implies $I_{\ell{i}}'\tilde{\xi}_i'>I_{\ell{i}}\tilde{\xi}_i$
and completes the proof.\hfill{$\blacksquare$}

By Lemma \ref{lemma:implicit_func_monotonicity} and the monotonicity
of the logarithmic function, it can be seen that
$\tilde{R}_i(\{I_{ki}\}_{k{\ne}i})$ is nonincreasing w.r.t. $I_{ki}$
for all $k{\ne}i$. We prove the convexity of
$\tilde{R}_i(\{I_{ki}\}_k)$ w.r.t. $I_{ki}$, $k{\ne}i$ by showing
that $\partial\tilde{R}_i(\{I_{ki}\}_k)/\partial{I_{ki}}$ is
nondecreasing w.r.t. $I_{ki}$ for all $k{\ne}i$, i.e.,
$\partial^2\tilde{R}_i(\{I_{ki}\}_k)/\partial{I_{ki}}^2\ge0$ for all
$k{\ne}i$. Let $\ell\in\{1,\dots,i-1,i+1,\dots,K\}$. By
\eqref{eq:partial_derivative}, we can explicitly express
$\partial\tilde{R}_i(\{I_{ki}\}_k)/\partial{I_{ki}}$ as
\eqref{eq:partial_derivative_tilde} on the top of the next page.
\begin{figure*}[t]
\begin{align}
&\frac{\partial\tilde{R}_i(\{I_{ki}\}_k)}{\partial{I_{\ell{i}}}}\notag\\
&~=\frac{-I_{ii}\tilde{\xi}_i(\{I_{ki}\}_{k{\ne}i})}{\ln2\cdot(1+\tilde{\xi}_i(\{I_{ki}\}_{k{\ne}i})I_{ii})}\left[(1+I_{\ell{i}}\tilde{\xi}_i(\{I_{ki}\}_{k{\ne}i}))\cdot\left(\sigma_i^2+\sum_{j\ne{i}}\frac{I_{ji}}{1+I_{ji}\tilde{\xi}_i(\{I_{ki}\}_{k{\ne}i})}\right)\right]^{-1}\notag\\
&~=\frac{-I_{ii}\tilde{\xi}_i(\{I_{ki}\}_{k{\ne}i})}{\ln2\cdot(1+I_{ii}\tilde{\xi}_i(\{I_{ki}\}_{k{\ne}i}))}\left[(1+I_{\ell{i}}\tilde{\xi}_i(\{I_{ki}\}_{k{\ne}i}))\cdot\left(\sigma_i^2+\displaystyle{\sum_{j\ne{i},j\ne\ell}}\frac{I_{ji}}{1+I_{ji}\tilde{\xi}_i(\{I_{ki}\}_{k{\ne}i})}\right)+I_{\ell{i}}\right]^{-1}.\label{eq:partial_derivative_tilde}
\end{align}
\hrulefill\vspace{-.3cm}
\end{figure*}
By Lemma \ref{lemma:implicit_func_monotonicity}, we can see that
$I_{ii}\tilde{\xi}_i(\{I_{ki}\}_{k\ne{i}})/(1+I_{ii}\tilde{\xi}_i(\{I_{ki}\}_{k\ne{i}}))\ge0$
is nonincreasing w.r.t. $I_{\ell{i}}$ while
$(1+I_{ji}\xi_i(\{I_{ki}\}_{k\ne{i}}))^{-1}$ is nondecreasing and
$(1+I_{\ell{i}}\xi_i(\{I_{ki}\}_{k\ne{i}}))$ is strictly increasing
w.r.t. $I_{\ell{i}}$. Therefore,
$\partial\tilde{R}_i(\{I_{ki}\}_{k{\ne}i})/\partial{I_{\ell{i}}}$ is
nondecreasing w.r.t. $I_{\ell{i}}$, and hence
$\tilde{R}_i(\{I_{ki}\}_{k{\ne}i})$ is convex w.r.t. $I_{ki}$,
$\forall{k{\ne}i}$.\hfill{$\blacksquare$}

\section{Monotonic Optimization by Polyblock Outer Approximation
Algorithm}\label{sec:intr_MonoOpt_POA}

Monotonic optimization refers to maximizing a nondecreasing function
over an intersection of so called \textit{normal sets}
\cite{Tuy2000_Monotonic}. By definition, a nonnegative set
$\mathcal{D}\subseteq\mathbb{R}_+$ is called normal if for any two
points $\mathbf{d}_1\succeq\mathbf{d}_2\succeq\zerob$,
$\mathbf{d}_1\in\mathcal{D}$ implies $\mathbf{d}_2\in\mathcal{D}$.
Let $f:\mathbb{R}^N\to\mathbb{R}$ be a nondecreasing function and
$\mathcal{D}\subseteq\mathbb{R}_+^N$ be a compact normal set. Then,
the monotonic optimization problem can be formulated as
\begin{equation}\label{monotonic_opt}
\max_{\xb\in\mathcal{D}}~f(\xb).
\end{equation}
According to \cite{Tuy2000_Monotonic}, this class of problems can be
optimally solved by a POA algorithm which is briefly reviewed in
this section. Before presenting the POA algorithm, some essential
definitions are given as follows.
\begin{Definition}\label{def:polyblock}
A set is called a polyblock if it is the union of a finite number of
boxes, where a box associated with a vertex $\vb\in\mathbb{R}_+^N$
is referred to the hyperrectangle
$\mathcal{B}(\vb)=\{\xb\in\mathbb{R}_+^N\mid\zerob\preceq\xb\preceq\vb\}$.
\end{Definition}

\begin{Definition}\label{def:proper_vertex}
A vertex $\vb\in\mathcal{P}$ is called a proper vertex of the
polyblock $\mathcal{P}$ if there is no vertex $\vb'\in\mathcal{P}$
such that $\vb'\succeq\vb$ and $\vb'\ne\vb$.
\end{Definition}

The main effort of the POA algorithm lies in constructing a sequence
of polyblocks $\{\mathcal{P}[0],\mathcal{P}[1],\dots\}$ such that
\begin{subequations}\label{eq:outer_approximation}
\begin{align}
&\mathcal{P}[0]\supseteq\mathcal{P}[1]\supseteq\cdots\supseteq\mathcal{P}[n]\supseteq\cdots\supseteq\mathcal{D},\label{eq:outer_approximation_a}\\
&\lim_{n\to\infty}\max_{\xb\in\mathcal{P}[n]}f(\xb)=\max_{\xb\in\mathcal{D}}f(\xb).\label{eq:outer_approximation_b}
\end{align}
\end{subequations}
In general, the initial polyblock can simply be a single box
associated with a vertex $\vb^\star[0]$, i.e.,
$\mathcal{P}[0]=\mathcal{B}(\vb^\star[0])$, such that
$\mathcal{D}\subseteq \mathcal{B}(\vb^\star[0])$. Given the
polyblock $\mathcal{P}[n-1]$ at the $(n-1)$th iteration, the
polyblock $\mathcal{P}[n]$ for iteration $n$ can be constructed as
follows. Let $\mathcal{V}[n-1]$ denote the set of proper vertices of
$\mathcal{P}[n-1]$. Firstly, we find a point $\vb^\star[n-1] \in
\mathcal{P}[n-1]$ that maximizes $f(\xb)$ over $\mathcal{V}[n-1]$,
and hence maximizes $f(\xb)$ over $\mathcal{P}[n-1]$ according to
the monotonicity of $f(\xb)$. Specifically, we find $\vb^\star[n-1]$
such that
\begin{equation}\label{eq:POA_outer_best}
\vb^\star[n-1]\in\mathrm{arg}\max_{\vb\in\mathcal{V}[n-1]}~f(\vb)\subseteq\mathrm{arg}\max_{\xb\in\mathcal{P}[n-1]}~f(\xb).
\end{equation}
Problem \eqref{eq:POA_outer_best} can be solved by enumerating all
the vertexes in $\mathcal{V}[n-1]$. Secondly, we search for the
intersection of the right-upper boundary of $\mathcal{D}$ and the
ray from the origin to $\vb^\star[n-1]$, i.e.,
\begin{equation}\label{eq:intersection}
\tilde{\vb}[n-1]=\beta^\star[n-1]\vb^\star[n-1],~~\beta^\star[n-1]=\mathrm{arg}\max_{\beta\vb^\star[n-1]\in\mathcal{D}}\beta.
\end{equation}
Problem \eqref{eq:intersection} can be solved by bisecting over $\beta$, which entails checking the feasibility of $\beta\vb^\star[n-1]\in\mathcal{D} $ iteratively. Thirdly, using $\vb^\star[n-1]$ and $\tilde{\vb}[n-1]$, we generate $N$ new vertices by
\begin{equation}\label{eq:new_vertices}
\vb[n,i]=\vb^\star[n-1]-(v_i^\star[n-1]-\tilde{v}_i[n-1])\eb_i,~i=1,\dots,N,
\end{equation}
where $v_i^\star[n-1]$ and $\tilde{v}_i[n-1]$  are the
$i$th element of $\vb^\star[n-1]$ and $\tilde{\vb}[n-1]$, respectively, and $\eb_i$ is the $i$th
column of the $N{\times}N$ identity matrix. Then, a
new vertex set $\mathcal{V}[n]$ is obtained as
\begin{align}\label{eq:Vn}
\mathcal{V}[n]\!=\!\mathcal{V}[n\!-\!1]\bigcup\{\vb[n,1],\dots,\vb[n,N]\}\backslash\{\vb^\star[n\!-\!1]\},
\end{align}
which leads to a new polyblock for the $n$th iteration
\begin{align}
\mathcal{P}[n]=\bigcup_{\vb\in \mathcal{V}[n]} \mathcal{B}(\vb).
\end{align}

Notice that $\mathcal{P}[n]\subseteq\mathcal{P}[n-1]$ since
$\vb^\star[n-1]\succeq\vb[n,i]$ for all $i=1,\dots,N$. Besides, by
\eqref{eq:intersection} and by the fact that $\mathcal{D}$ is
normal, one can infer that the intersection of $\mathcal{D}$ and
$\mathcal{P}[n-1]\backslash\mathcal{P}[n]$ must be empty\footnote{A
brief proof is as follows. From \eqref{eq:Vn}, we see that
$\mathcal{P}[n-1]\backslash\mathcal{P}[n]=\{\xb\mid\tilde{\vb}[n-1]\prec\xb\preceq\vb^\star[n-1]\}$.
If the intersection of $\mathcal{D}$ and
$\mathcal{P}[n-1]\backslash\mathcal{P}[n]$ is not empty. Then there
must exist a point $\bar{\xb}$ such that $\bar{\xb}\in\mathcal{D}$
and $\tilde{\vb}[n-1]\prec\bar{\xb}\preceq\vb^\star[n-1]$. This
implies that there exists $\beta\in(\beta^\star[n-1],1]$ such that
$\tilde{\xb}=\beta\vb^\star[n-1]\preceq\bar{\xb}$ and
$\tilde{\xb}\in\mathcal{D}$ (since $\mathcal{D}$ is normal), which
however contradicts with the optimality of $\beta^\star[n-1]$ to
problem \eqref{eq:intersection}.}, implying that
$\mathcal{D}\in\mathcal{P}[n]$.
As a result, the polyblocks
$\{\mathcal{P}[0],\mathcal{P}[1],\dots,\mathcal{P}[n],\dots\}$
generated in this manner indeed satisfy
\eqref{eq:outer_approximation_a}. {In addition, it has been shown in
\cite[Theorem 1]{Tuy2000_Monotonic} that
\eqref{eq:outer_approximation_b} also holds true.} Thus, by
\eqref{eq:POA_outer_best}, the sequence
$\{f(\vb^\star[n])\}_{n=0}^\infty$ monotonically converges to the
optimal value of problem \eqref{monotonic_opt} from above. On the
other hand, let
$\bar{\vb}[n]=\mathrm{arg}{\displaystyle\max_{\vb\in\{\bar{\vb}[n-1],\tilde{\vb}[n]\}}}~f(\vb)$,
where $\tilde{\vb}[n]\in\mathcal{D}$ for all $n\ge0$. Then the
sequence $\{f(\bar{\vb}[n])\}_{n=0}^\infty$ will also monotonically
converge to the optimal value of problem \eqref{monotonic_opt} from
below \cite[Theorem 1]{Tuy2000_Monotonic}. Therefore, the gap
between $f(\vb^\star[n])$ and $f(\bar{\vb}[n])$ can be used as an
estimate of the difference between $f(\bar{\vb}[n])$ and the optimal
value of \eqref{monotonic_opt}, serving as a stopping criterion for
the POA algorithm. Finally, the POA algorithm for problem
\eqref{monotonic_opt} is summarized in Algorithm \ref{alg:poa}.
%

\begin{algorithm}[h]
\caption{POA algorithm for solving problem \eqref{monotonic_opt}}
\begin{algorithmic}[1]\label{alg:poa}
\STATE {\bf Initialization:} Set the solution accuracy as
$\delta\ge0$, and set $n:=0$.

\STATE Set $\mathcal{V}[0]:=\{\vb^\star[0]\}$, where $\vb^\star[0]$
can be any vector such that $\mathcal{D} \subseteq \mathcal{B}(\vb^\star[0])$;

\STATE Compute $\tilde{\vb}[0]$ by \eqref{eq:intersection}, and set
$\bar{\vb}[0]:=\tilde{\vb}[0]$;

\WHILE{$f(\vb^\star[n])-f(\bar{\vb}[n])>\delta$}

\STATE $n:=n+1$;

\STATE Set
$\mathcal{V}[n]=\mathcal{V}[n-1]\bigcup\{\vb[n,1],\dots,\vb[n,N]\}\backslash\{\vb^\star[n-1]\}$,
where $\vb[n,i]$, $i=1,\dots,N$, are given by
\eqref{eq:new_vertices};

\STATE Compute $\vb^\star[n]$ and $\tilde{\vb}[n]$ by
\eqref{eq:POA_outer_best} and \eqref{eq:intersection}, respectively;

\STATE Set
$\bar{\vb}[n]:=\mathrm{arg}{\displaystyle\max_{\vb\in\{\bar{\vb}[n-1],\tilde{\vb}[n]\}}}~f(\vb)$;

\ENDWHILE

\STATE {\bf Output} $f(\vb^\star[n])$ as the approximation of the
optimal value of \eqref{monotonic_opt}.
\end{algorithmic}
\end{algorithm}

\section*{Acknowledgement}

The authors would like to thank Prof. Che Lin of National Tsing Hua
University, Hsinchu, Taiwan, for valuable discussions in preparing
this manuscript.

\bibliographystyle{IEEEtran}
\bibliography{references}

\begin{thebibliography}{10}
\providecommand{\url}[1]{#1}
\csname url@samestyle\endcsname
\providecommand{\newblock}{\relax}
\providecommand{\bibinfo}[2]{#2}
\providecommand{\BIBentrySTDinterwordspacing}{\spaceskip=0pt\relax}
\providecommand{\BIBentryALTinterwordstretchfactor}{4}
\providecommand{\BIBentryALTinterwordspacing}{\spaceskip=\fontdimen2\font plus
\BIBentryALTinterwordstretchfactor\fontdimen3\font minus
  \fontdimen4\font\relax}
\providecommand{\BIBforeignlanguage}[2]{{%
\expandafter\ifx\csname l@#1\endcsname\relax
\typeout{** WARNING: IEEEtran.bst: No hyphenation pattern has been}%
\typeout{** loaded for the language `#1'. Using the pattern for}%
\typeout{** the default language instead.}%
\else
\language=\csname l@#1\endcsname
\fi
#2}}
\providecommand{\BIBdecl}{\relax}
\BIBdecl

\bibitem{Li2013}
W.-C. Li, T.-H. Chang, C.~Lin, and C.-Y. Chi, ``Outage constrained weighted sum
  rate maximization for {MISO} interference channel by pricing-based
  optimization,'' in \emph{Proc. 2013 IEEE ICASSP}, Vancouver, BC, May 26-31,
  2013, pp. 4799--4803.

\bibitem{Lee_2012_ComMag}
J.~Lee, Y.~Kim, H.~Lee, B.~L. Ng, D.~Mazzarese, J.~Liu, W.~Xiao, and Y.~Zhou,
  ``Coordinated multipoint transmission and reception in {LTE}-advanced
  systems,'' \emph{IEEE Commun. Mag.}, vol.~50, no.~11, pp. 44--50, Nov. 2012.

\bibitem{Gesbert10JSAC}
D.~Gesbert, S.~Hanly, H.~Huang, S.~S. Shitz, O.~Simeone, and W.~Yu,
  ``Multi-cell {MIMO} cooperative networks: A new look at interference,''
  \emph{IEEE J. Sel. Areas Commun.}, vol.~28, no.~9, pp. 1380--1408, Dec. 2010.

\bibitem{Bjornson11TSP}
E.~Bj\"{o}rnson, N.~J. Jald\'{e}n, M.~Bengtsson, and B.~Ottersten, ``Optimality
  properties, distributed strategies, and measurement-based evaluation of
  coordinated multicell {OFDMA} transmission,'' \emph{IEEE Trans. Signal
  Process.}, vol.~59, no.~12, pp. 6086--6101, Dec. 2011.

\bibitem{Annapureddy2011}
V.~S. Annapureddy and V.~V. Veeravalli, ``Sum capacity of {MIMO} interference
  channels in the low interference regime,'' \emph{IEEE Trans. Inf. Theory},
  vol.~57, no.~5, pp. 2565--2581, May 2011.

\bibitem{Liu_11}
Y.-F. Liu, Y.-H. Dai, and Z.-Q. Luo, ``Coordinated beamforming for {MISO}
  interference channel: {Complexity} analysis and efficient algorithms,''
  \emph{IEEE Trans. Signal Process.}, vol.~59, no.~3, pp. 1142--1157, Mar.
  2011.

\bibitem{Cai_2012}
D.~Cai, T.~Quek, C.~W. Tan, and S.~Low, ``Max-min {SINR} coordinated multipoint
  downlink transmission--duality and algorithms,'' \emph{IEEE Trans. Signal
  Process.}, vol.~60, no.~10, pp. 5384--5395, Oct. 2012.

\bibitem{Zakhour_09}
R.~Zakhour and D.~Gesbert, ``Coordination on the {MISO} interference channel
  using the virtual {SINR} framework,'' in \emph{Proc. Int. ITG Workshop on
  Smart Antennas}, Berlin, Germany, Feb. 16-18, 2009.

\bibitem{Zhang_Cui_2010}
R.~Zhang and S.~Cui, ``Cooperative interference management with {MISO}
  beamforming,'' \emph{IEEE Trans. Signal Process.}, vol.~58, pp. 5450--5458,
  Oct. 2010.

\bibitem{Kim_2011}
S.-J. Kim and G.~Giannakis, ``Optimal resource allocation for {MIMO} ad hoc
  cognitive radio networks,'' \emph{IEEE Trans. Inf. Theory}, vol.~57, pp.
  3117--3131, May 2011.

\bibitem{Shi2011_IteMMSE}
Q.~Shi, M.~Razaviyayn, Z.-Q. Luo, and C.~He, ``An iteratively weighted {MMSE}
  approach to distributed sum-utility maximization for a {MIMO} interfering
  broadcast channel,'' \emph{IEEE Trans. Signal Process.}, vol.~59, no.~9, pp.
  4331--4340, Sep. 2011.

\bibitem{Nguyen_11}
D.~H.~N. Nguyen and T.~Le-Ngoc, ``Multiuser downlink beamforming in multicell
  wireless systems: {A} game theoretical approach,'' \emph{IEEE Trans. Signal
  Process.}, vol.~59, pp. 3326--3338, July 2011.

\bibitem{Weeraddana_2013}
P.~Weeraddana, M.~Codreanu, M.~Latva-aho, and A.~Ephremides, ``Multicell {MISO}
  downlink weighted sum-rate maximization: {A} distributed approach,''
  \emph{IEEE Trans. Signal Process.}, vol.~61, no.~3, pp. 556--570, 2013.

\bibitem{Hong2012}
M.-Y. Hong and Z.-Q. Luo, ``Signal processing and optimal resource allocation
  for the interference channel,'' \emph{Academic Press Library in Signal
  Process., arXiv:1206.5144v1}, 2013.

\bibitem{Larsson_08}
E.~G. Larsson and E.~A. Jorswieck, ``Competition versus cooperation on the
  {MISO} interference channel,'' \emph{IEEE J. Sel. Areas Commun.}, vol.~26,
  no.~7, pp. 1059--1069, Sep. 2008.

\bibitem{Larsson_etal2009_mag}
E.~G. Larsson, E.~A. Jorswieck, J.~Lindblom, and R.~Mochaourab, ``Game theory
  and the flat-fading {Gaussian} interference channel,'' \emph{IEEE Signal
  Process. Mag.}, vol.~26, no.~5, pp. 18--27, Sep. 2009.

\bibitem{Schmidt_09}
D.~A. Schmidt, C.~Shi, R.~A. Berry, M.~L. Honig, and W.~Utschick, ``Distributed
  resource allocation schemes: {Pricing} algorithms for power control and
  beamformer design in interference networks,'' \emph{IEEE Signal Process.
  Mag.}, vol.~26, no.~5, pp. 53--63, Sep. 2009.

\bibitem{Jorswieck2010_POA}
E.~A. Jorswieck and E.~G. Larsson, ``Monotonic optimization framework for the
  two-user {MISO} interference channel,'' \emph{IEEE Trans. Commun.}, vol.~58,
  no.~7, pp. 2159--2168, July 2010.

\bibitem{Utschick2012_POA}
W.~Utschick and J.~Brehmer, ``Monotonic optimization framework for coordinated
  beamforming in multicell networks,'' \emph{IEEE Trans. Signal Process.},
  vol.~60, no.~4, pp. 1899--1909, Apr. 2012.

\bibitem{Zhang_2012_POA}
L.~Liu, R.~Zhang, and K.-C. Chua, ``Achieving global optimality for weighted
  sum-rate maximization in the {$K$}-user {Gaussian} interference channel with
  multiple antennas,'' \emph{IEEE Trans. Wireless Commun.}, vol.~11, no.~5, pp.
  1933--1945, May 2012.

\bibitem{Kandukuri02}
S.~Kandukuri and S.~Boyd, ``Optimal power control in interference-limited
  fading wireless channels with outage-probability specifications,'' \emph{IEEE
  Trans. Wireless Commun.}, vol.~1, pp. 46--55, Jan. 2002.

\bibitem{Tan_2011}
C.~W. Tan, ``Optimal power control in {Rayleigh}-fading heterogeneous
  networks,'' in \emph{Proc. IEEE INFOCOM}, Shanghai, April 10-15, 2011, pp.
  2552--2560.

\bibitem{Huang_2012_GLOBECOM}
Y.~Huang, C.~W. Tan, and B.~Rao, ``Outage balancing in multiuser {MISO}
  networks: {Network} duality and algorithms,'' in \emph{Proc. IEEE GLOBECOM},
  Anaheim, CA, Dec. 3-7, 2012, pp. 3918--3923.

\bibitem{Ghosh_10}
S.~Ghosh, B.~D. Rao, and J.~R. Zeidler, ``Outage-efficient strategies for
  multiuser {MIMO} networks with channel distribution information,'' \emph{IEEE
  Trans. Signal Process.}, vol.~58, pp. 6312--6324, Dec. 2010.

\bibitem{Lindblom_11}
J.~Lindblom, E.~Karipidis, and E.~G. Larsson, ``Outage rate regions for the
  {MISO} interference channel: {Definitions} and interpretations,''
  \emph{http://arxiv.org/abs/1106.5615v1}.

\bibitem{Park2012}
J.~Park, Y.~Sung, D.~Kim, and H.~V. Poor, ``Outage probability and outage-based
  robust beamforming for {MIMO} interference channels with imperfect channel
  state information,'' \emph{IEEE Trans. Wireless Commun.}, vol.~11, pp.
  3561--3573, June 2012.

\bibitem{Li_13}
W.-C. Li, T.-H. Chang, C.~Lin, and C.-Y. Chi, ``Coordinated beamforming for
  multiuser {MISO} interference channel under rate outage constraints,''
  \emph{IEEE Trans. Signal Process.}, vol.~61, pp. 1087--1103, Mar. 2013.

\bibitem{Li_14_cplx}
\BIBentryALTinterwordspacing
W.-C. Li, T.-H. Chang, and C.-Y. Chi, ``Multicell coordinated beamforming with
  rate outage constraint--{Part I}: {Complexity} analysis,'' \emph{submitted to
  IEEE Trans. Signal Process.}, 2014. [Online]. Available:
  \url{http://arxiv.org/abs/1405.2982}
\BIBentrySTDinterwordspacing

\bibitem{Razaviyayn_2013BSUM}
M.~Razaviyayn, M.~Hong, and Z.-Q. Luo, ``A unified convergence analysis of
  block successive minimization methods for nonsmooth optimization,''
  \emph{SIAM Journal on Optimization}, vol.~23, no.~2, pp. 1126--1153, 2013.

\bibitem{Tuy2000_Monotonic}
H.~Tuy, ``Monotonic optimization: {Problems} and solution approaches,''
  \emph{SIAM J. Optimization}, vol.~11, no.~2, pp. 464--494, 2000.

\bibitem{Mo2000}
J.~Mo and J.~Walrand, ``Fair end-to-end window-based congestion control,''
  \emph{IEEE/ACM Trans. Networking}, vol.~8, pp. 556--567, Oct. 2000.

\bibitem{BK:Bertsekas1999}
D.~P. Bertsekas, \emph{Nonlinear Programming}, 2nd~ed.\hskip 1em plus 0.5em
  minus 0.4em\relax Belmont, MA: Athena Scientific, 1999.

\bibitem{Krantz_Parks02}
S.~G. Krantz and H.~R. Parks, \emph{The Implicit Function Theorem: History,
  Theory, and Applications}.\hskip 1em plus 0.5em minus 0.4em\relax Boston, MA:
  Birkh\"{a}user, 2002.

\bibitem{BK:BoydV04}
S.~Boyd and L.~Vandenberghe, \emph{Convex {O}ptimization}.\hskip 1em plus 0.5em
  minus 0.4em\relax Cambridge, UK: Cambridge University Press, 2004.

\bibitem{Marks1978}
B.~R. Marks and G.~P. Wright, ``A general inner approximation algorithm for
  nonconvex mathematical programs,'' \emph{Operations Research}, vol.~26, pp.
  681--683, 1978.

\bibitem{Bjornson2012_BRB}
E.~Bj\"{o}rnson, G.~Zheng, M.~Bengtsson, and B.~Ottersten, ``Robust monotonic
  optimization framework for multicell {MISO} systems,'' \emph{IEEE Trans.
  Signal Process.}, vol.~60, pp. 2508 --2523, May 2012.

\bibitem{Luo2010_SPM}
Z.-Q. Luo, W.-K. Ma, A.~M.-C. So, Y.~Ye, and S.~Zhang, ``Semidefinite
  relaxation of quadratic optimization problems,'' \emph{IEEE Signal Process.
  Mag.}, vol.~27, pp. 20 --34, May 2010.

\bibitem{cvx}
M.~Grant and S.~Boyd, ``{CVX}: Matlab software for disciplined convex
  programming, version 1.21,'' \url{http://cvxr.com/cvx}, Apr. 2011.

\end{thebibliography}

\end{document}